\documentclass[aps,prd,twocolumn,showpacs,floats,floatfix,letterpaper,nofootinbib,superscriptaddress,]{revtex4-2}

\usepackage{hyperref}
\usepackage{amssymb,amsmath,latexsym,mathrsfs}
\usepackage{graphicx,subfigure}
\usepackage{epsfig}
\usepackage{varioref,xr-hyper}
\usepackage{color}
\usepackage{multirow}
\usepackage{array}
\hypersetup{
	colorlinks   = true,
	citecolor    = blue
}

\begin{document}

\title{Cosmic backreaction and the mean redshift drift from symbolic regression}
	
\author{S. M. Koksbang} 
\email{koksbang@cp3.sdu.dk}
\affiliation{CP3-Origins, University of Southern Denmark, Campusvej 55, DK-5230 Odense M, Denmark}

\begin{abstract}
The possibility of obtaining symbolic expressions for cosmic backreaction is explored through a case study of so-called 2-region models. By using the publicly available symbolic regression algorithm AI Feynman, it is shown that the kinematical backreaction from a single 2-region model can be well described as a function of the mean redshift (or, equivalently, the volume averaged scale factor). A single expression depending on the redshift/scale factor as well as a model parameter, $f$, that can accurately describe the backreaction for a significant range of models is naturally more complicated but is also achieved with percent-level accuracy.
\newline\indent
Data sets of redshift drift in the 2-region models are also considered. Again utilizing AI Feynman, expressions for the redshift drift are found. In particular, an expression for the difference between the mean redshift drift and the drift of the mean redshift in terms of the kinematical backreaction is easily obtained for a single 2-region model. An accurate symbolic expression that describes this difference for an array of 2-region models is achieved by using the redshift as a feature instead of the kinematical backreaction.
\end{abstract}
	
\maketitle
	
\section{Introduction}
Since the first cosmological solution to Einstein's equation was presented in 1917 \cite{Einstein} (see e.g. also \cite{history_Einstein}), modern cosmology has been based on the cosmological principle: The notion that the Universe is (statistically) homogeneous and isotropic on large scales. More importantly, modern cosmology has since its birth been based on the (often implicit) assumption that the cosmological principle implies that one can insert an exactly spatially homogeneous content and corresponding metric assumptions into Einstein's equation when wishing to describe the Universe. This means that standard cosmology is based on the Friedmann-Lemaitre-Robertson-Walker (FLRW) models. Questioning this procedure of using the FLRW models to interpret observations and describe the dynamics of the Universe is a cornerstone in the research field of \emph{inhomogeneous cosmology}. Note that two separate issues must be addressed. One issue concerns the effect inhomogeneities have on observations. Some effects are well-known and even make up an important part of standard cosmology. This is for instance the case for fluctuations in the cosmic microwave background. But in inhomogeneous cosmology the focus is also on the possibility that \emph{mean} observations may deviate from the observational relations given by an FLRW model meant to describe the large-scale spatial average of the Universe. This has been studied for decades, with a variety of different approaches such as using inhomogeneous cosmological models \cite{CliftonBull, Kantowski, Early2region-ish, Tzavara, Tzavara2, biswas, Marra, Marra2, tetradis, Wasserman, Clitfon_void, Wessel, Kostov, flanagan, flanagan2, Fleury, Fluery2, Bolejko, Bolejko2, Bolejk03, szybka, mattsson, Clifton_weird, Clifton_Wheeler, lattice, tardis, syksyCMB, selv_model, selv_model2}, numerical cosmology \cite{Fleury_num, Starkman, hayley, selv_num, selv_num2, adamek1, adamek2, adamek3, simsilun} and analytical considerations \cite{Fleury_ana, Fleury_ana2, Fleury_ana3, Fleury_ana4, Fleury_ana5, Bolejko_ana, Bolejko_ana2, misinterp, discreteDM, nearFLRW, light1, light2, Linder1, Linder2} including different spacetime slicings \cite{lightcone, lightcone_2, lightcone_3, lightcone_jacobi, lightcone_average, lightcone_backreaction, Buchert_lightcone}. In addition to this, there is the separate issue of whether or not the large-scale/spatially averaged behavior of the Universe indeed \emph{does} follow FLRW dynamics. This was studied as early as in \cite{first_av}, but today the most popular way of studying the average evolution of an inhomogeneous spacetime is through the Buchert formalism \cite{fluid1, fluid2, fluid3} developed decades later. There is good sense in using this formalism since the work presented in \cite{light1, light2} combined with that in especially \cite{another_look, Hellaby} shows that spatial averages based on the Buchert formalism can be directly related to observations if 1) averages are made on spatial hypersurfaces of statistical homogeneity and isotropy, 2) mean observations are based on averaging over several, random light rays sampling spacetime fairly, i.e. without avoiding certain regions such as e.g. overdensities, and 3) structures evolve slowly compared to the time it takes a light ray to traverse the homogeneity scale (assumed to exist).
\newline\indent
With the Buchert averaging formalism, averages of scalars are computed as volume weighted averages, i.e. as
\begin{align}
	s_D := \frac{\int_D s dV}{\int_D dV} = \frac{\int_D s dV}{V},
\end{align}
where $s$ is a scalar being averaged over a spatial domain $D$, and $dV$ is the proper (Riemannian) infinitesimal spatial volume element. When using this simplest form of the formalism, it is assumed that spacetime is foliated with spatial hypersurfaces orthogonal to the fluid flow and that the lapse function (the time-time component of the metric tensor) is set to 1. This is assumed throughout. Note that this foliation requires that there is no vorticity. The foliation corresponds to assuming that the line element of the spacetime can be written as
\begin{align}
	ds^2 = -dt^2 + g_{ij}dx^i dx^j,
\end{align}
where $i,j \in[1,2,3]$ are used for indicating spatial indexes. Greek letters will be used to denote spacetime indexes running over $0,1,2,3$.
\newline\indent
Applying the Buchert formalism to the Hamiltonian constraint and the Raychaudhuri equation leads to ($c= 1$ throughout)
\begin{align}\label{eq:Buchert}
3H_D^2 & = 8\pi G \rho_D - \frac{1}{2}R_D - \frac{1}{2}Q\\
3\frac{\ddot a_D}{a_D} & = -4\pi G\rho_D + Q.
\end{align}
A possible cosmological constant is omitted from the equations as it will not be included in the models considered here. The spatially averaged Hubble parameter, $H_D$, is defined as $H_D:=\frac{\dot a_D}{a_D}$, where $a_D:=(V/V_0)^{1/3}$ is the volume averaged scale factor normalized to 1 at present time. Present time evaluation is indicated by a subscripted zero and $V$ is the proper volume of the averaging domain $D$. Dots are used to indicate partial derivatives with respect to the time coordinate. $R_D$ is the (spatially averaged) spatial curvature scalar and $Q:=2/3\left[\left(\theta^2 \right)_D -\left(\theta_D \right)^2  \right] - \left(\sigma_{\mu\nu}\sigma^{\mu\nu} \right)_D  $ is known as the kinematical backreaction, computed through averages of the fluid expansion scalar, $\theta$, and its shear tensor, $\sigma_{\mu\nu}$. Note that subscripted $D$'s are used to denote that quantities are either averages themselves or given in terms of average quantities (such as $a_D$ given in terms of the averaging volume). It would also be appropriate to add such a subscript to the kinematical backreaction since this quantity is given in terms of spatial averages and certainly depends on the choice of averaging domain $D$. However, in order to simplify the notation and reduce subscript-clutter in the following sections, $Q$ is not given the subscript $D$, but it is here stressed that $Q$ does represent an averaged quantity and indeed has no local counterpart. Note lastly that $\rho_D\propto a_D^{-3}$, equivalent to the FLRW limit\cite{fluid1}.
\newline\indent
The Buchert equations are very similar to the Friedmann equations which govern the dynamics of the FLRW models, with the key differences being 1) that the averaged spatial curvature does not have to be proportional to the inverse squared averaged scale factor, and 2) the extra term, $Q$, which vanished identically in the FLRW limit. The two components $R_D$ and $Q$ thus make up the {\em cosmic backreaction}. They depend on each other through the  the so-called integrability condition
\begin{align}\label{eq:constraint}
	a_D^{-6} \left(a_D^6 Q \right)^.  + a_D^{-2} \left(a_D^2 R_D  \right)^. = 0,
\end{align}
which ensures that the two previous equations are consistent with each other.
\newline\newline
A major obstacle in \emph{inhomogeneous cosmology} is determining the dependence of $Q$ and $R_D$ on $a_D$. A small number of studies based on different forms of perturbation theory as well as numerical simulations have given some minor indication that the relationship may be very simple, in the form of $Q\propto a_D^{\pm 1}$ \cite{1overa1,1overa2,1overa3}. In addition, it was early noted \cite{scaling} that there is a very simple set of solutions to the integrability condition, namely the scaling solutions which take the form
\begin{align}
R_D & = R_{D_0} a_D^n\\
Q &= -\frac{n+2}{n+6}R_D,
\end{align}
with $n\neq -6,-2$. These solutions have been used for studying observational effects of backreaction in e.g. \cite{n1_temp1, n1_temp2, template, GW_n}. However, aside from the minor indications just mentioned that $Q$ may in some instances scale as $\propto a_D^{\pm 1}$, there is little physical justification for the scaling solution which is instead mainly a result of mathematical convenience and simplicity. Additional physical justification can be attributed the relation $Q\propto a_D^{-1}$ since this represents the leading large-scale mode \cite{1overa3}. Overall, this means that the scaling solutions can mainly be used for proof-of-principle studies such as in \cite{n1_temp1} where the scaling solution was used to show that backreaction can in principle explain the Hubble tension \cite{tension1, tension2}. In order to seriously constrain backreaction through observations, more information is needed regarding the dependence of $Q$ on $a_D$ and possibly other quantities such as the average matter density etc. The same must be obtained for $R_D$ but note that since $R_D$ and $Q$ are related by the integrability condition (as well as the top line in equation \ref{eq:Buchert}, i.e. the first Buchert equation), we can in principle obtain one, once we have the other.
\newline\newline
This article and its accompanying letter \cite{accompanying_letter} represent the first step towards learning about the parameterization of cosmic backreaction by using machine learning. Specifically, results from an initial investigation into the possibility and use of determining the parameterization of $Q$ and $R_D$ in terms of $a_D$ with symbolic regression are presented.
\newline\indent
Symbolic regression is an automated regression analysis where the algorithms learn symbolic expressions that accurately describe a given set of data (see e.g.  \cite{ESR,AIFeynman_1}). With $Q$ and $R_D$ parameterized in terms of e.g. the volume averaged scale factor, it becomes possible to constrain backreaction with redshift-distance relations since $Q$ and $R_D$ enter into the redshift distance relation given according to \cite{light1}
\begin{align}\label{eq:av_DA}
	H_D\frac{d}{d\left\langle z\right\rangle }\left( (1+\left\langle z\right\rangle )H_D\frac{d\left\langle D_A\right\rangle }{d\left\langle z\right\rangle } \right) = -4\pi G\rho_D \left\langle D_A\right\rangle ,
\end{align}
where $z, D_A$ are the redshift and angular diameter distance and triangular brackets are used to denote mean values, i.e. mean relations obtained by averaging over several random lines of sight and ensuring that the requirements 1)-3) mentioned earlier are fulfilled (see e.g. \cite{light1} for a detailed discussion of the requirements). Note that $1+\left\langle z\right\rangle  = 1/a_D$ \cite{light1} and that $Q$ and $R_D$ enter into the equation through $H_D$.
\newline\indent
While the redshift-distance relation can be fairly easily related to spatial averages as discussed above, this is not true for all types of observations. One observable that does not seem to be easily described by spatial averages is the redshift drift. The redshift drift is the change in the observed redshift of a source due to cosmic expansion. It was first discussed in \cite{Sandage,McVittie}. As first demonstrated in \cite{another_look} and later corroborated in \cite{Hellaby}, the mean redshift drift is not as simply related to spatially averaged quantities as the redshift-distance relation is. This has further been studied in \cite{Asta_dz1, Asta_dz2} which support the result: The mean redshift drift is not in general equal to the drift of the mean redshift in an inhomogeneous spacetime.
\newline\indent
Besides studying the parameterization of cosmic backreaction it will in the following be studied to what extent it is possible to identify a symbolic expression for the redshift drift in terms of spatially averaged quantities. This possibility is interesting for (at least) two reasons. First of all, if an analytic expression relating the mean redshift drift to spatial averages can be found, it becomes possible to use redshift drift to constrain spatially averaged quantities. This is not currently possible unless one is in the FLRW limit. Secondly, a symbolic expression could be useful as a guide to analytical/theoretical studies of redshift drift and e.g. drive an investigation into the theoretical underpinnings of such a symbolic expression obtained through machine learning. In a similar vain, parameterizing $Q$ and $R_D$ through $a_D$ not only makes it possible to constrain these using observables, but the resulting symbolic expression can advise analytical studies into this parameterization with the goal to e.g. learn about under what conditions non-negligible backreaction occurs.
\newline\newline
In this first study, a simple toy-model will be used to obtain data sets for the task of learning symbolic expressions for the kinematical backreaction, spatially averaged curvature, and mean redshift drift. Specifically, 2-region models will be used. These models were chosen because they are fast and simple to use for generating backreaction and redshift drift data. In section \ref{sec:2region}, 2-region models are introduced together with the method used for computing  redshift drift in these models. Results from symbolic regression and a feature importance analysis are then presented in section \ref{sec:results} before a discussion and concluding remarks are given in section \ref{sec:conclusion}.

\section{2-region models}\label{sec:2region}
A 2-region model is a toy-model representing an inhomogeneous universe constructed as a disjoint ensemble of two different specific FLRW solutions to Einstein's equation. As the individual FLRW solutions are disjoint, the 2-region models are \emph{not} exact solutions to Einstein's equation, but they are nonetheless useful for initial investigations and proof-of-principle studies; the models are fairly easy to construct and are fairly \emph{un}demanding computationally. This type of model was introduced in \cite{2region_first1, 2region_first2}, where backreaction was considered in 2-region models in a spacetime represented by the disjoint ensemble of two FLRW regions. The models were later generalized to consist of an ensemble of multiple versions of each of the two distinct FLRW solutions, with the individual regions being joined sequentially along light rays to mimic a statistically homogeneous and isotropic universe \cite{2region_light, another_look, 2region_observations}. This is the version of the 2-region models which will be considered here. Note that the 2-region models can be viewed as a simple version of the multi-scale models of \cite{multiscale1,multiscale2} and a simple version of the model presented in \cite{simple_timescape}.
\newline\indent
The considered 2-region models will be constructed as the ensemble of an empty FLRW model and matter+curvature FLRW model, where the latter is modeled to have positive curvature. In this case, the scale factors of the two different types of regions can be related according to (see e.g. \cite{2region_first1,2region_first2} for details)
\begin{align}
	t& = t_0\frac{\phi-\sin(\phi)}{\phi_0-\sin(\phi_0)}\\
	a_u& = \frac{f_u^{1/3}}{\pi}(\phi-\sin(\phi))\\
	a_o & = \frac{f_o^{1/3}}{2}(1-\cos(\phi)),
\end{align}
where $a_u$ is the local scale factor of the underdense (empty) FLRW region and $a_o$ is the local scale factor of the overdense region. The parameter $\phi$ is a parameter that is used for convenience when describing matter+curvature FLRW regions and is sometimes called the development angle. The relative fractions, $f_u,f_o$, of the two region types in the total ensemble at $\phi = \pi$ are related by $f_u = 1-f_o$. Following the original work in \cite{2region_first1, 2region_first2}, present time is set to be at $\phi_0 = 3/2\pi$. There is an additional free parameter which can be set by noting that $H_D = H_u(1-v+vh)$, where $v:=a_o^3/(a_o^3+a_u^3)$ and $h = H_o/H_u$. Since $H_u = 1/t$ we see that $t_0 = (1-v_0+v_0h_0)/H_{D_0}$ so we need to fix either $t_0$ or $H_{D_0}$. Here, the choice $H_{D_0} = 70$km/s/Mpc is made.
\newline\newline
In the following, $f_o$ is used as a free parameter and will for notational simplicity be referred to simply as $f$. Note that $f$ is a constant but that this does {\em not} mean that the volume fraction of over- and underdense regions is constant in the considered models. $f$ is merely constant because it represents the volume fraction {\em at a specific time}, namely the time corresponding to $\phi = \pi $.

\subsection{Redshift drift in 2-region models}
The redshift drift, $\delta z$, along light rays propagating through consecutive FLRW regions can be computed  using the equations (see \cite{another_look})
\begin{align}
	\frac{dt}{dr} &= -a\\
	\frac{dz}{dr}& = (1+z)\dot a\\
	\frac{d\delta z}{dr} &= \dot a\delta z + (1+z)\ddot a\delta t\\
	\frac{d\delta t}{dr} &=-\dot a	\delta t,
\end{align}
where the scale factor is always evaluated as the local value. The parameter $\delta t$ represents the difference in emission time of the two signals of the redshift. Typical values for $\delta t_0$ are in the range 10-30 years based on the expectation regarding how long observation periods will be used with upcoming surveys (see e.g. \cite{dz_to_5} for an example). The choice $\delta t_0 = 30$ was made for all the results presented below, but it is noted that this choice is of little importance here since $\delta t_0$ largely just gives an overall scaling of $\delta z$. This is true even for inhomogeneous models as long as $\delta t_0$ is not chosen to be very large compared to the typical dynamical scale of the inhomogeneities, at least in the models studied here where light paths are all repeatable (see e.g. \cite{repeatable} regarding repeatable light paths).
\newline\newline
When solving the above equations, a choice must be made regarding how big the individual FLRW regions should be along the light rays. As mentioned in \cite{another_look} the resulting redshift drift does not depend significantly on the exact choice as long as the regions are not excessively large ($\lesssim 1$Gpc), in agreement with condition 3) from the introduction. Here, inhomogeneities are chosen to be of order $10-100$Mpc at present time. As also found in \cite{another_look}, the spatial position (underdense or overdense region) of the present-time observer is of little significance for the overall redshift drift signal. The observer is therefore simply always positioned in the overdense region in the following (an arbitrary choice). Since the redshift drift depends only minimally on the observer position and structures sizes, the redshift drift computed along a single light ray will be a good approximation to the mean redshift drift.
\newline\newline
The data studied in the following was obtained by propagating single light rays through 2-region models. Data was only collected along a single light ray for each model since the results of \cite{another_look} indicate that very little difference is obtained by computing the mean of the redshift and redshift drift along different light rays with, say, differently placed (present-time) observers. In the following, the redshift and redshift drift along an individual light ray will therefore be used for approximating the mean values. To remind the reader that the generated data in reality represents data along a single light ray, triangular brackets will not be used around symbols ($z, \delta z$) representing the data.

\begin{figure*}
	\centering
	\subfigure[]{
		\includegraphics[scale = 0.5]{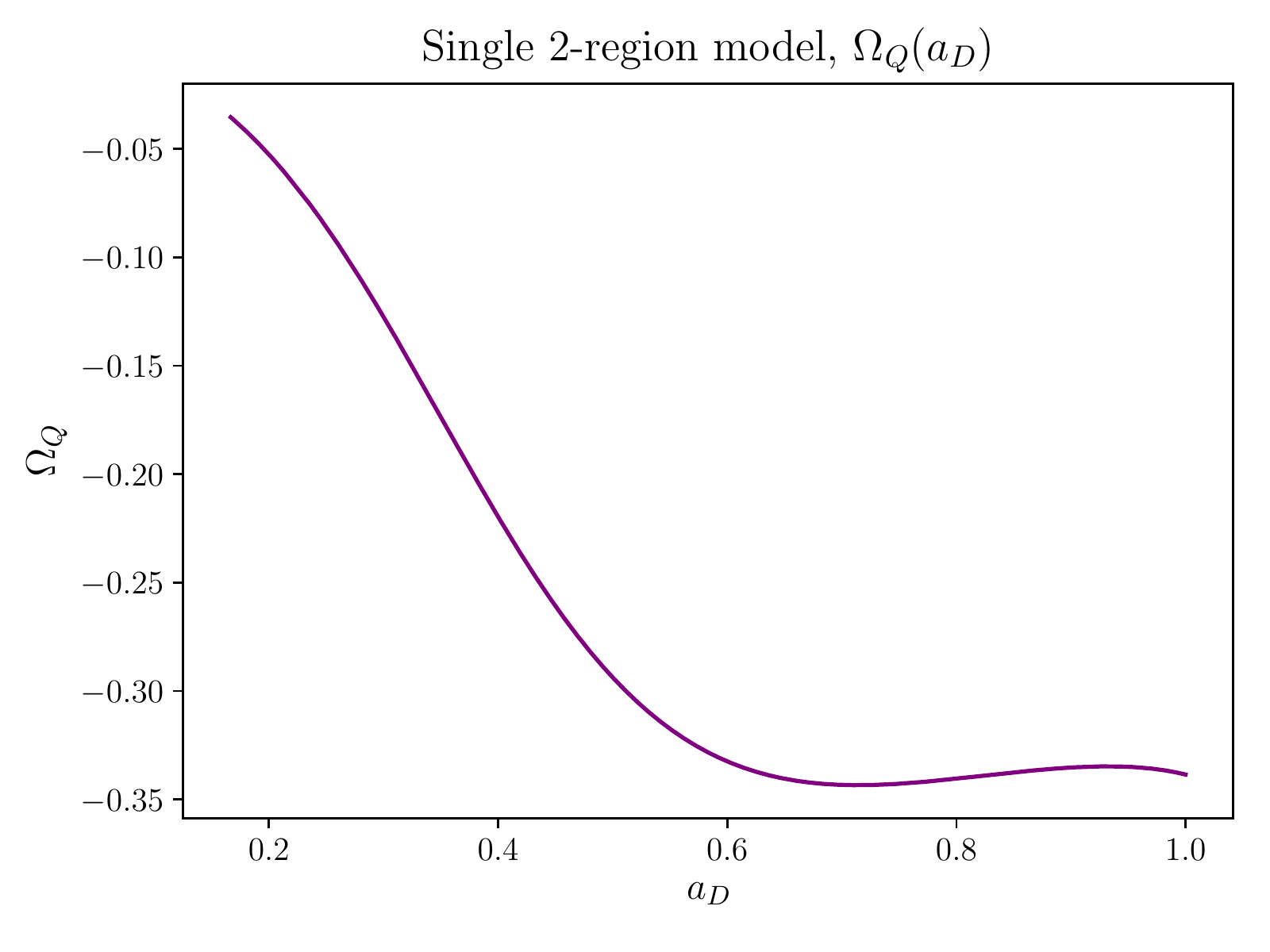}
	}
	\subfigure[]{
		\includegraphics[scale = 0.5]{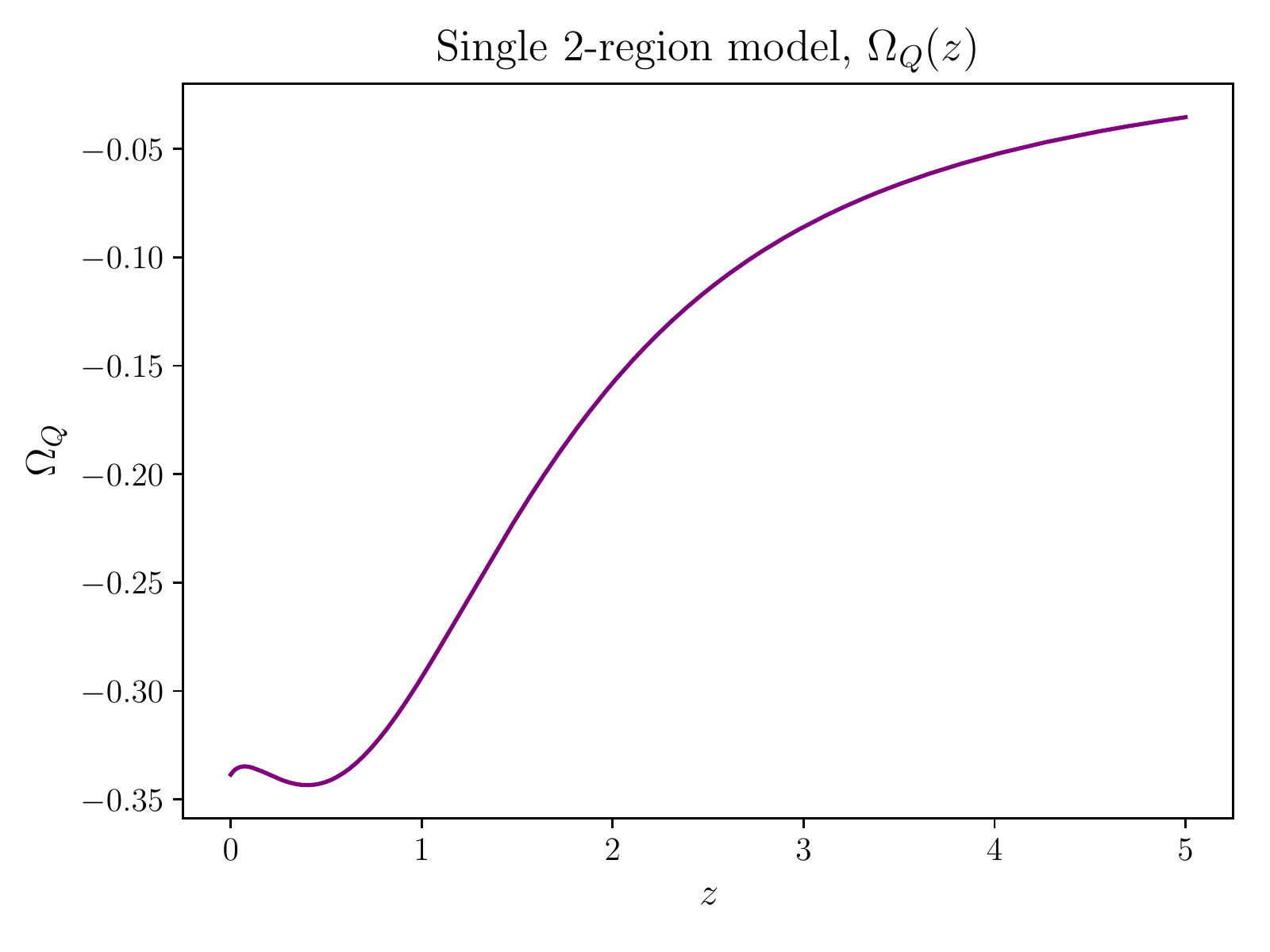}
	}\par
	\subfigure[]{
		\includegraphics[scale = 0.5]{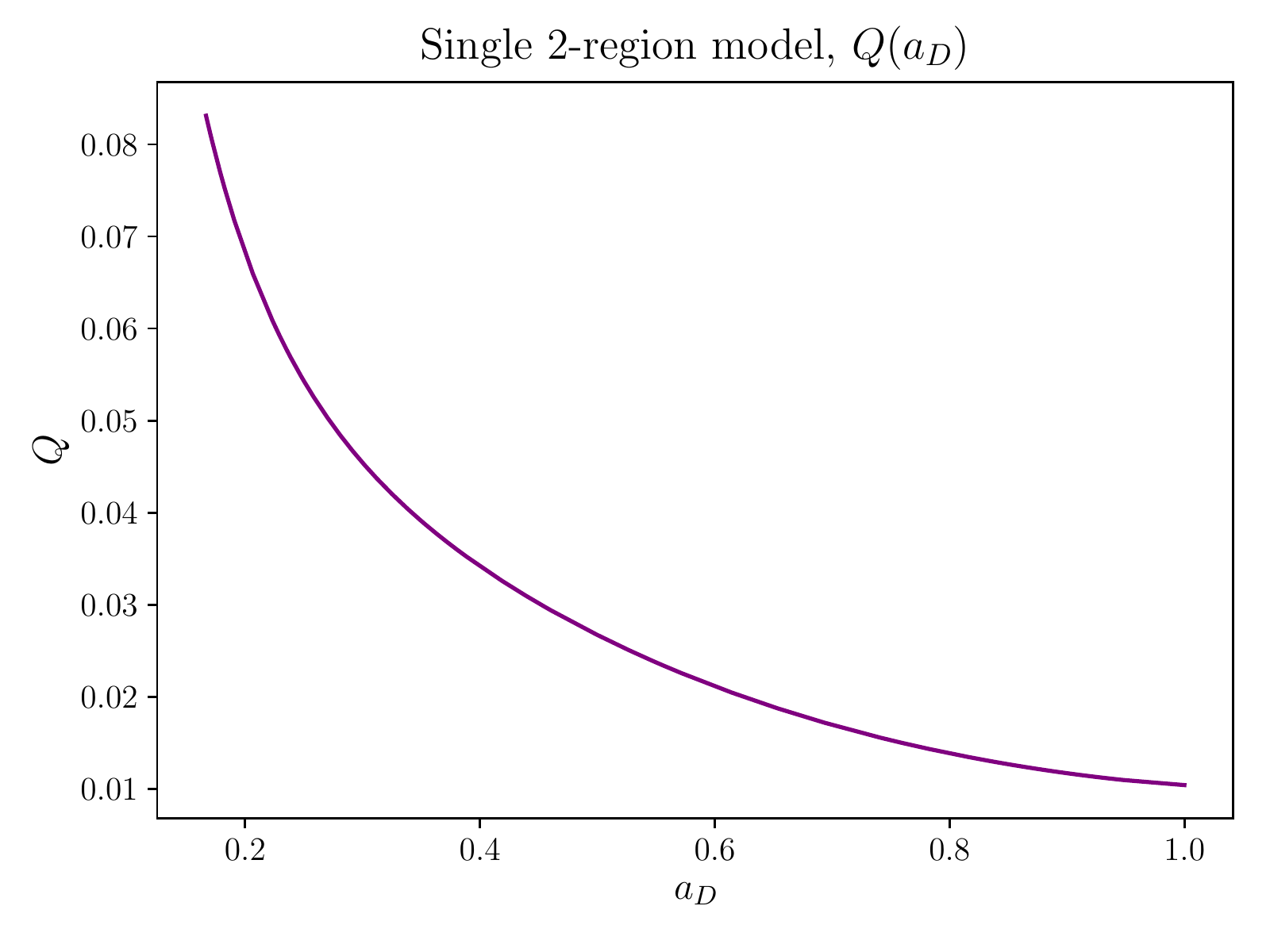}
	}
	\subfigure[]{
		\includegraphics[scale = 0.5]{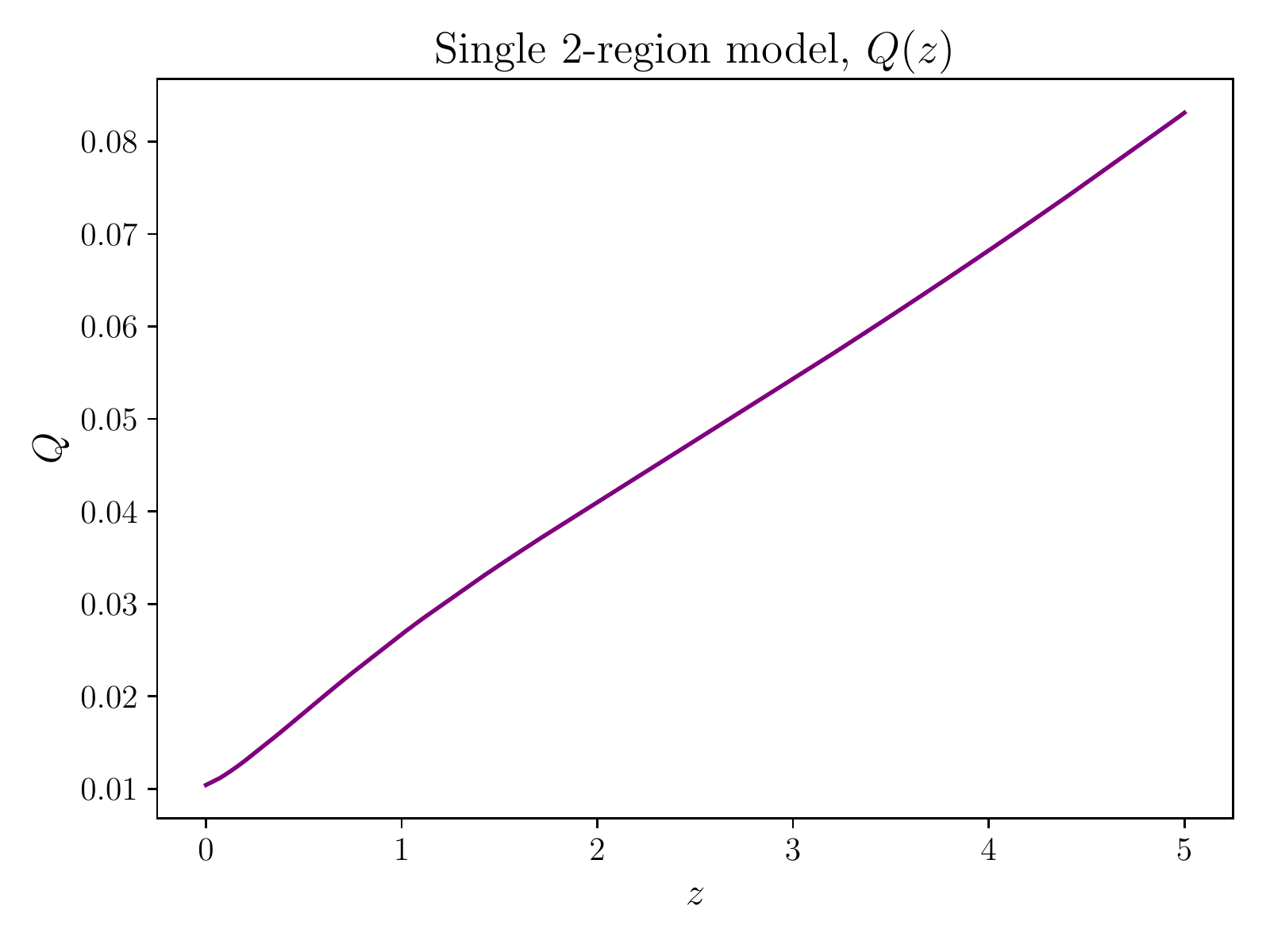}
	}
	\caption{The kinematical backreaction, $Q$, and its density parameter, $\Omega_Q$, as a function of $a_D$ and $z$.}
	\label{fig:Qza}
\end{figure*}
\section{Results}\label{sec:results}
There are several examples of publicly available software for doing symbolic regression such as PySR\footnote{https://github.com/MilesCranmer/PySR} and gplearn\footnote{https://gplearn.readthedocs.io/en/stable/} as well as algorithms specifically developed for (astro-)physics purposes, including e.g. ESR \cite{ESR} and AI Feynman \cite{AIFeynman_1, AIFeynman_2}. In the following, AI Feynman will be used because it was developed specifically with physics in mind and is publicly available in a ready-to-use format.
\newline\newline
AI Feynman fits analytical expressions to data sets of the form {\em (features, target)}, where the target is the target variable which depends on the different features supplied in the data array. For the study here, the targets are $Q$, $R_D$ and the redshift drift and features include mainly $z, a_D, f$. If we wish to obtain an analytical expression for, say, $Q$ in terms of $z,f$ we would generate data (a text file) with the values of $z, f$ together with the corresponding values of $Q$. In such a data file, $z$ and $f$ would be considered {\em features}, and $Q$ the {\em target}. Data files used here were generated with equidistant data points within the feature intervals given in the sections below.\\\\
Some of the data sets considered here have multiple features. When multiple features are introduced into the data sets, most machine learning algorithms including neural networks tend to perform better when trained on data sets with normalized features \cite{ML_bog}. In the data sets given to AI Feynman (which contains a neural network), the features could e.g. be normalized according to $x \rightarrow x_{\rm norm}:=(x-x_{\rm min})/(x_{\rm max}-x_{\rm min})$, where $x$ is a given feature value and $x_{\rm min}, x_{\rm max}$ are the minimum and maximum values, respectively, of the given feature in the data set. This is standard for many machine learning algorithms as it puts different features on equal footing with respect to the underlying algorithm. From a physical point of view this type of feature scaling is, however, unfortunate; it is much more useful to have an expression in terms of actual physical variables instead of scaled versions. Features scaled according to a normalization or standardization would make it difficult to use the resulting expressions since these could only be used with input parameters scaled appropriately. The data sets used here will therefore mainly be generated without feature scaling. When scaling is introduced in the following, this will be explicitly mentioned and explained.

\begin{figure*}
	\centering
	\subfigure[]{
		\includegraphics[scale = 0.5]{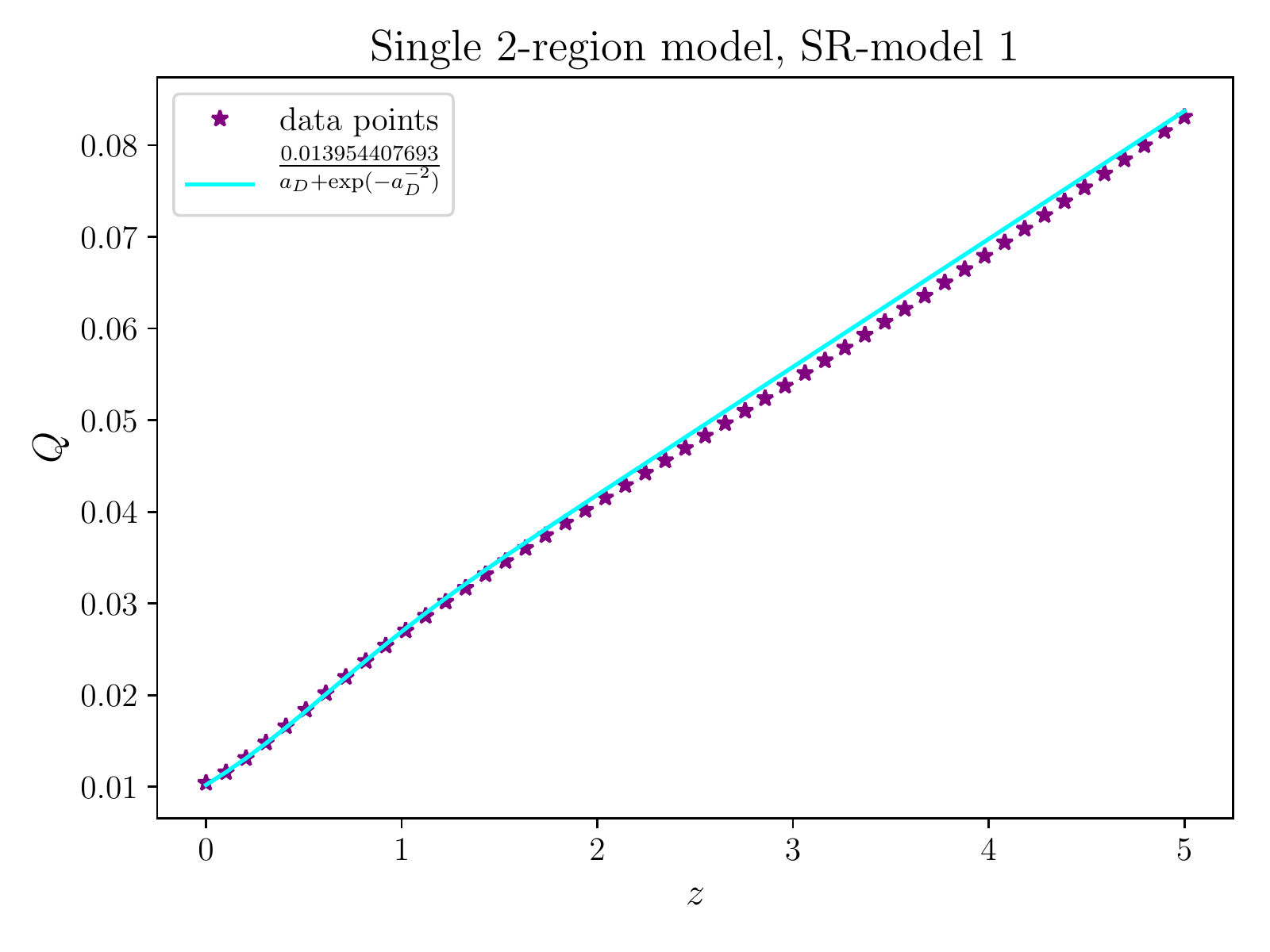}
	}
	\subfigure[]{
		\includegraphics[scale = 0.5]{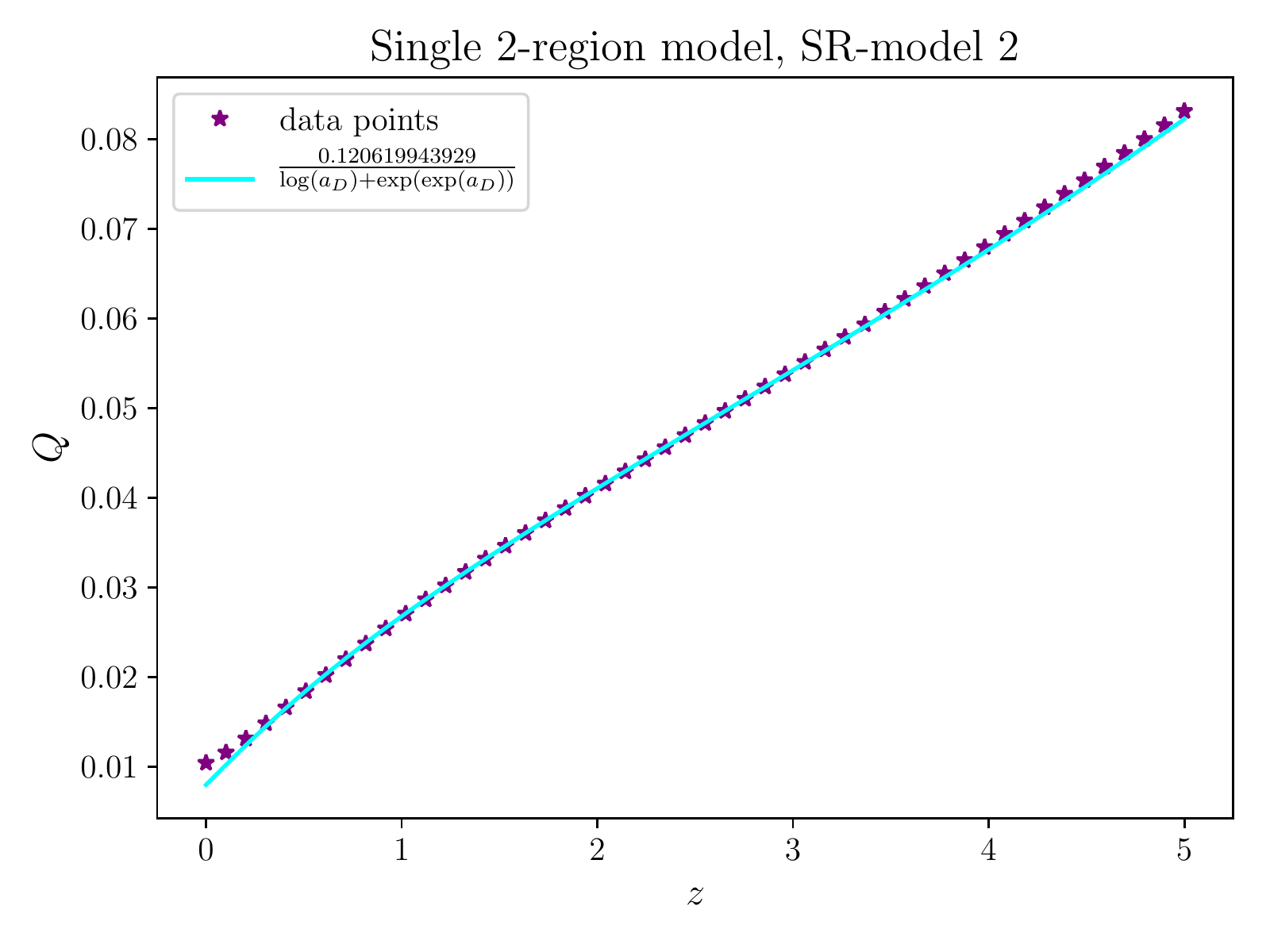}
	}\par
	\subfigure[]{
		\includegraphics[scale = 0.5]{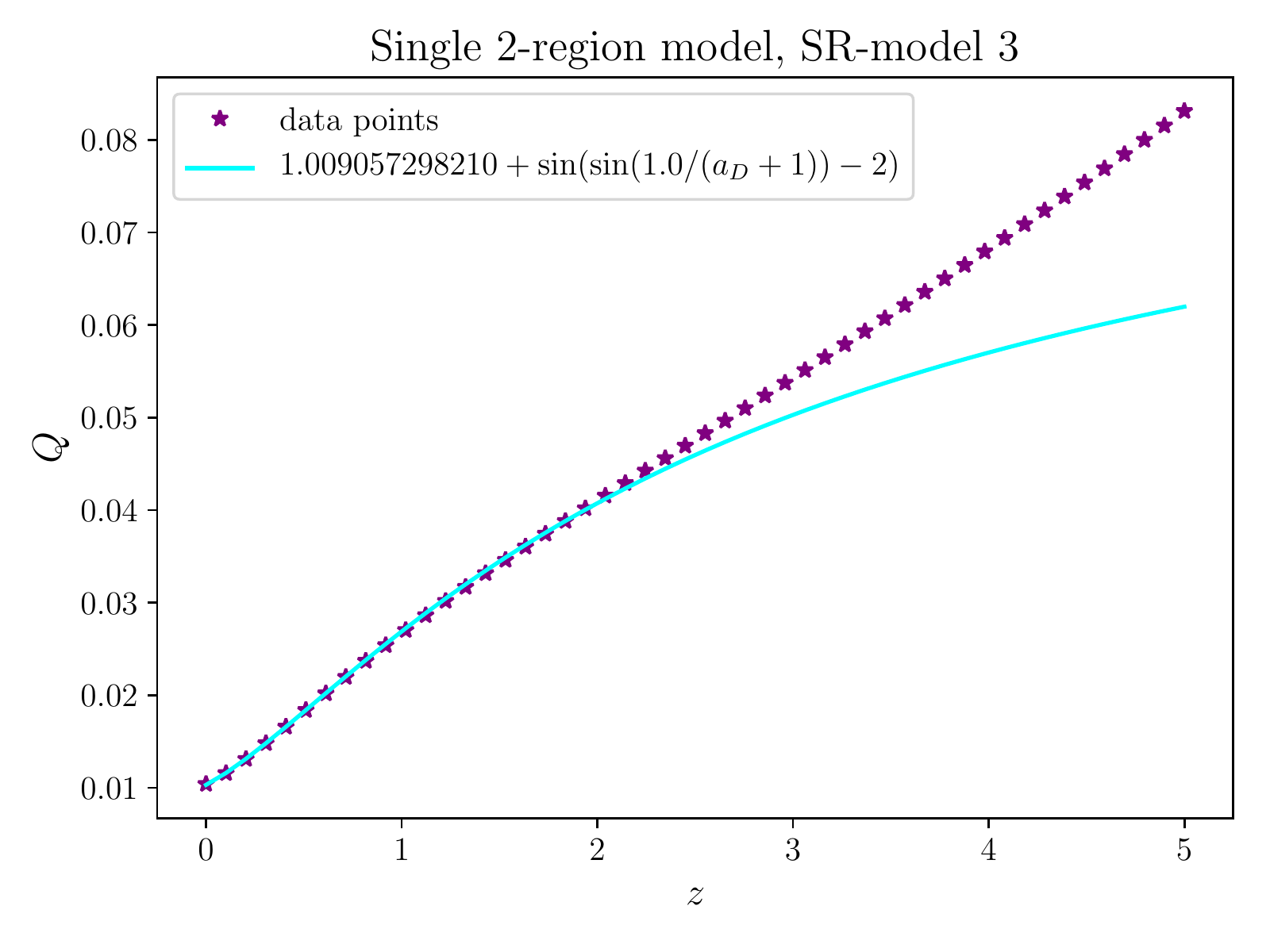}
	}
	\subfigure[]{
		\includegraphics[scale = 0.5]{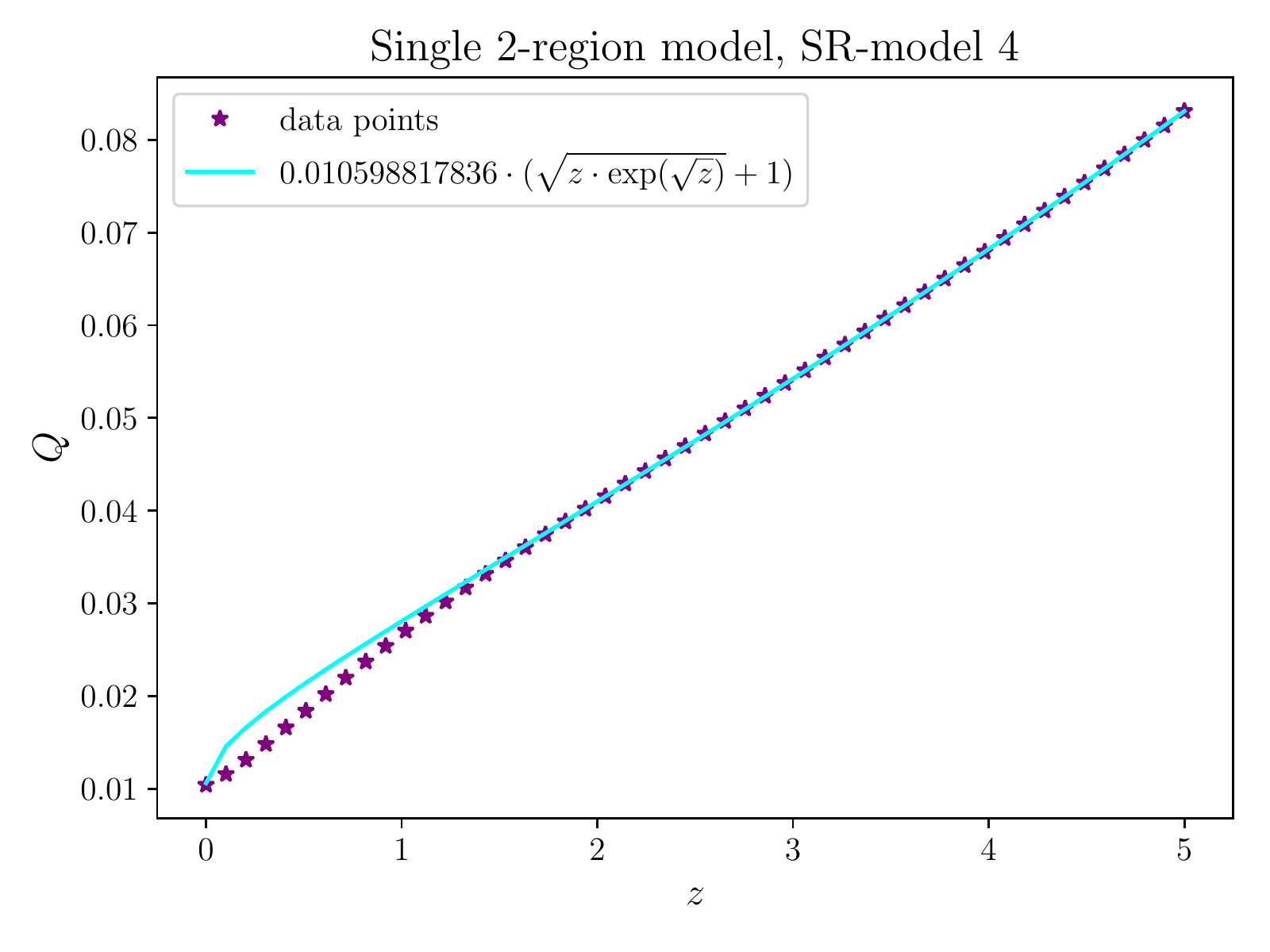}
	}
	\caption{Symbolic expressions for $(a_D,Q)$ and $(z,Q)$ found by AI Feynman plotted together with data points. If legends indicate a function in terms of $z$, the expression was obtained by presenting AI Feynman with data of the form $(z,Q)$, and equivalently for $a_D$.}
	\label{fig:FirstFit}
\end{figure*}
\subsection{Symbolic regression for a single 2-region model}\label{subsec:single}
As an initial test, AI Feynman is presented with backreaction data for a single specific 2-region model. Somewhat arbitrarily, the parameter $f = 0.25$ is used. This choice leads to a significant amount of average accelerated expansion and provides a kinematical backreaction that is fairly large but without the resulting model being wildly unrealistic with, for instance, large regions of overdensity and only small regions of underdensity at present time.
\newline\indent
As mentioned in the introduction, it is in principle not necessary to do symbolic regression on data of both $Q$ and $R_D$ since we may use the first Buchert equation or the integrability condition to obtain one, once we have the other. Therefore, for the  initial test presented in this subsection, the focus will be on the kinematical backreaction and the redshift drift. The reason for choosing $Q$ rather than $R_D$ is somewhat arbitrary with the main motivation being that $R_D$ has an FLRW counterpart while $Q$ does not. When modeling $R_D$ one must therefore also consider if e.g. subtracting an FLRW-part of $R_D$ is appropriate. No such considerations are necessary regarding $Q$.
\newline\newline
Backreaction data for this model has been generated with a different number of data points ranging between 50 and $50000$. AI Feynman was presented with the data set combinations $(z,Q)$, $(z,\Omega_{Q})$, $(a_D, Q)$ and $(a_D, \Omega_{Q})$, using all 4 available symbolic expressions files and with various choices of time limit for each brute force call, maximum degree of polynomial to be tried and number of epochs
\footnote{AI Feynman can be fed different types of information that restricts/directs the algorithm. For instance, AI Feynman comes with four different files of basic symbolic expressions it uses -- i.e. trigonometric functions, logarithms etc.. These can be modified or new files can be supplied by the user. For the work presented here, the four files were used without modifications and no new files were introduced. Another example of a parameter one can set to tune the algorithm is a time limit for the brute-force part of the AI Feynman algorithm. The time limits used here were in the interval 60-500 seconds. See e.g. the github page https://github.com/SJ001/AI-Feynman for details on the parameters/options that can be set.}.
The parameter $\Omega_{Q}$ is defined in analogy to the density parameters of the FLRW model, i.e. $\Omega_{Q}:= -Q/(6H_D^2)$. The redshift, $z$, used in these data sets was computed as $z = 1/a_D -1$ which, as mentioned in the introduction, is a good approximation of $\left\langle z\right\rangle $.
\newline\indent
\begin{figure}
	\centering
		\includegraphics[scale = 0.5]{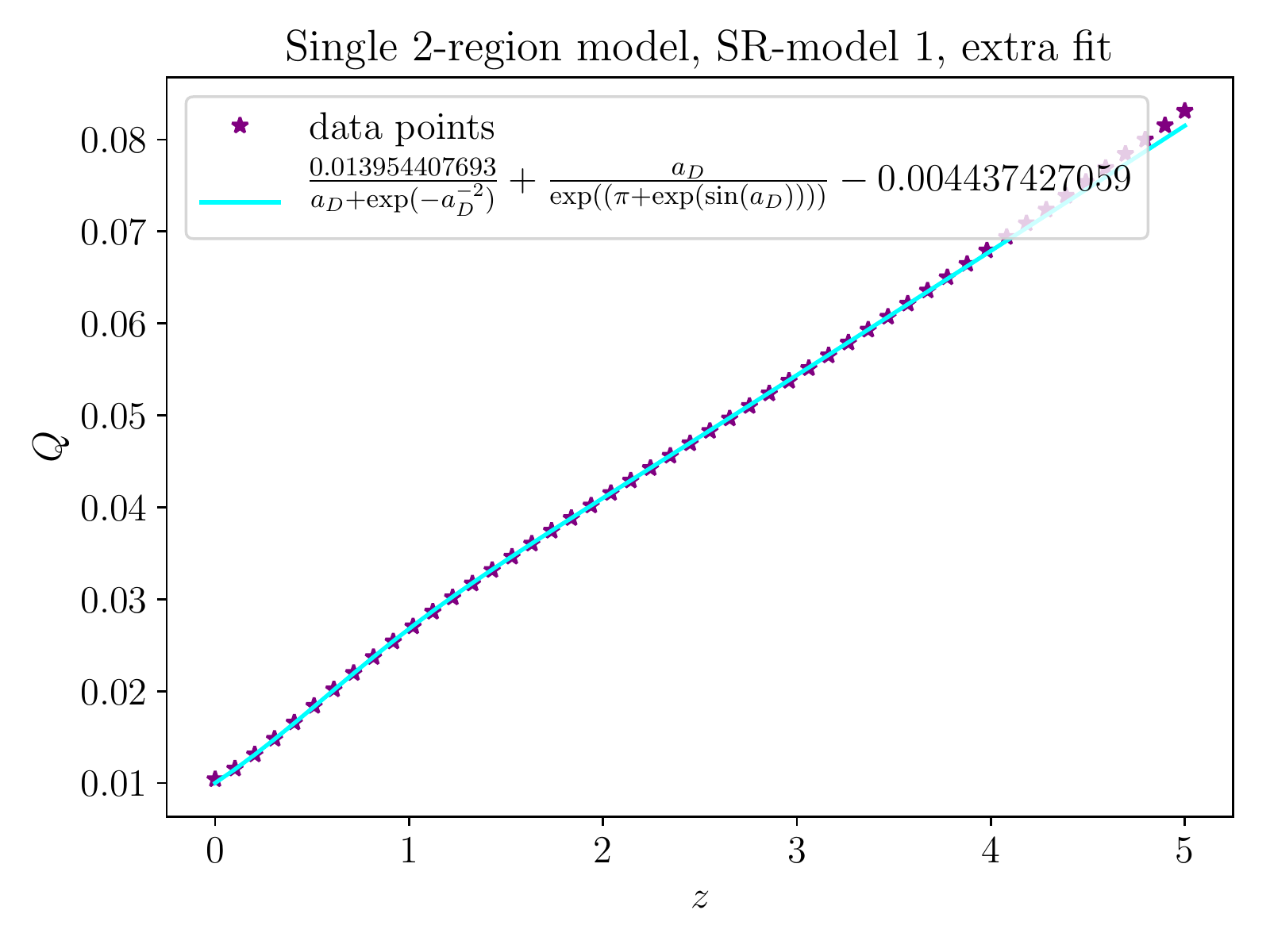}
	\caption{Symbolic expression for $(a_D,Q)$ found by AI Feynman used on data points of the type $(a_D, Q-\rm SRmodel1)$ plotted together with data points.}
	\label{fig:SecondFit}
\end{figure}
There still remains the choice of how large a parameter interval to use, i.e. how large the interval of $z$ (or $a_D$) should be. The larger an interval, the more difficult the regression task is. A larger interval will, however, represent a more general result which is typically desired. Here, the choice is made to consider the interval corresponding to $z\in[0,5]$. This choice is made based on the following considerations: Cosmic backreaction is expected to become important in the late universe, after nonlinear structures begin to form (see e.g. \cite{backreaction_latetime}), but at the same time, redshift drift is a quantity we expect to be able to observe all the way up to a redshift of $z=5$ (see e.g. \cite{dz_to_5}).
\newline\indent
Figure \ref{fig:Qza} shows $Q$ and $\Omega_{Q}$ plotted against both $z$ and $a_D$. By eye, it looks as though $Q$ has the simpler functional form, leading to the expectations that it will be easier to find an appropriate symbolic expression for $Q$ than for $\Omega_{Q}$ for the particular models studied here. The most accurate expressions found by AI Feynman were indeed obtained for $Q$, with a few of the more accurate expressions identified by AI Feynman shown in figure \ref{fig:FirstFit}. The legends indicate whether the data was generated as $(z,Q)$ or $(a_D,Q)$. Most of the expressions AI Feynman finds for the data only provide a reasonable fit to part of the data set. This is for instance the case for SR-model 3 in figure \ref{fig:FirstFit}, and to a lesser degree for SR-model 4. SR-models 1 and 2, however, are fairly accurate long the entire studied redshift interval. Specifically, the relative error between the data points and SR-model 1 is below 3\% for the entire redshift interval, while SR-model 2 has a sub-percent accuracy for $0.4\leq z\leq 4.4$ but becomes quite inaccurate at low redshifts with an error above 20\% for redshifts close to zero. On the redshift interval $z\in[4.4,5]$, the error is approximately 1\%, never reaching as much as $1.1\%$.
\newline\indent
There are several ways to proceed in order to obtain more accurate fits. For instance, the models depicted in figure \ref{fig:FirstFit} could be combined to form piece-wise expressions for $Q$ which are more accurate than any one of the expressions on the entire studied redshift interval. Indeed, by combining SR-model 2 and 3 one can obtain a piece-wise expression with sub-percent accuracy on almost the entire redshift interval and never exceeding $1.1\%$. This option can work well for symbolic expressions in one variable but quickly becomes inconvenient when considering expressions of multiple variables. Another option, which will briefly be studied here, is to use AI Feynman for a second iteration, now on data of the form $(a_D, Q-\rm SRmodelX)$, where SRmodelX is one of the models shown in figure \ref{fig:FirstFit}. This is a simple way to try to attempt to increase the accuracy of the symbolic expressions at the expense of the expressions becoming significantly more complex. Figure \ref{fig:SecondFit} shows the most accurate expression obtained by using SR-model1 for such an iterative procedure. The total expression is still not accurate to sub-percent precision on the entire studied redshift interval, but it {\em is} more accurate than SR-model1 alone, except at the data point with the smallest redshift value where the error has increased from 2\% to almost 4\% as well as for approximately $z\geq4.8$ where the error has increased from $1\%$ to around 2\%.
\newline\newline
We now move on to look at the redshift drift for the specified model. Again, the redshift drift could equally well be parameterized with $z$ or $a_D$, but there does not seem to be much to gain from using one rather than the other. Hence, we will only consider data points with $z$ as the feature (independent variable).
\newline\indent
\begin{figure*}
	\centering
	\subfigure[]{
		\includegraphics[scale = 0.5]{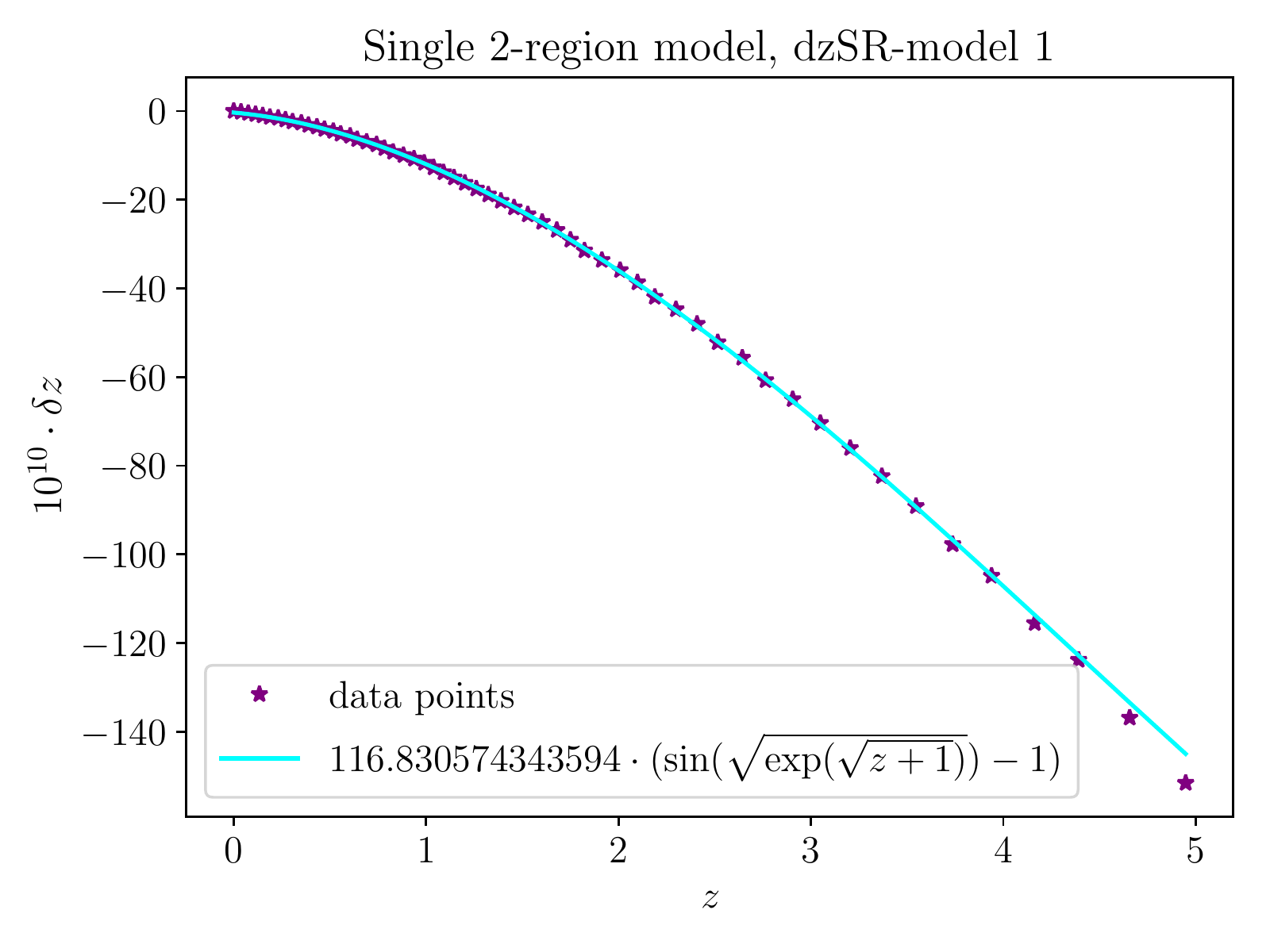}
	}
	\subfigure[]{
		\includegraphics[scale = 0.5]{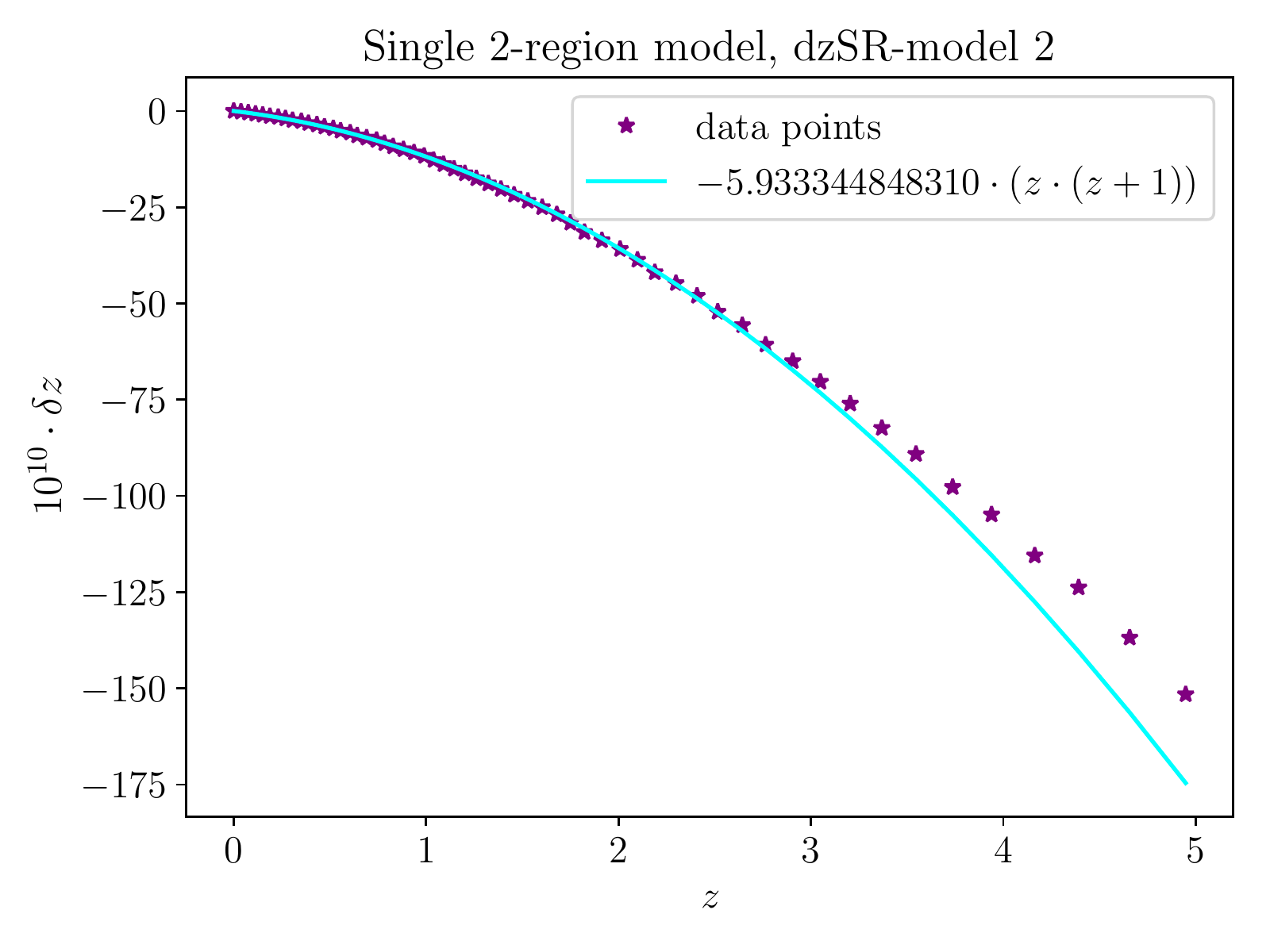}
	}\par
	\subfigure[]{
		\includegraphics[scale = 0.5]{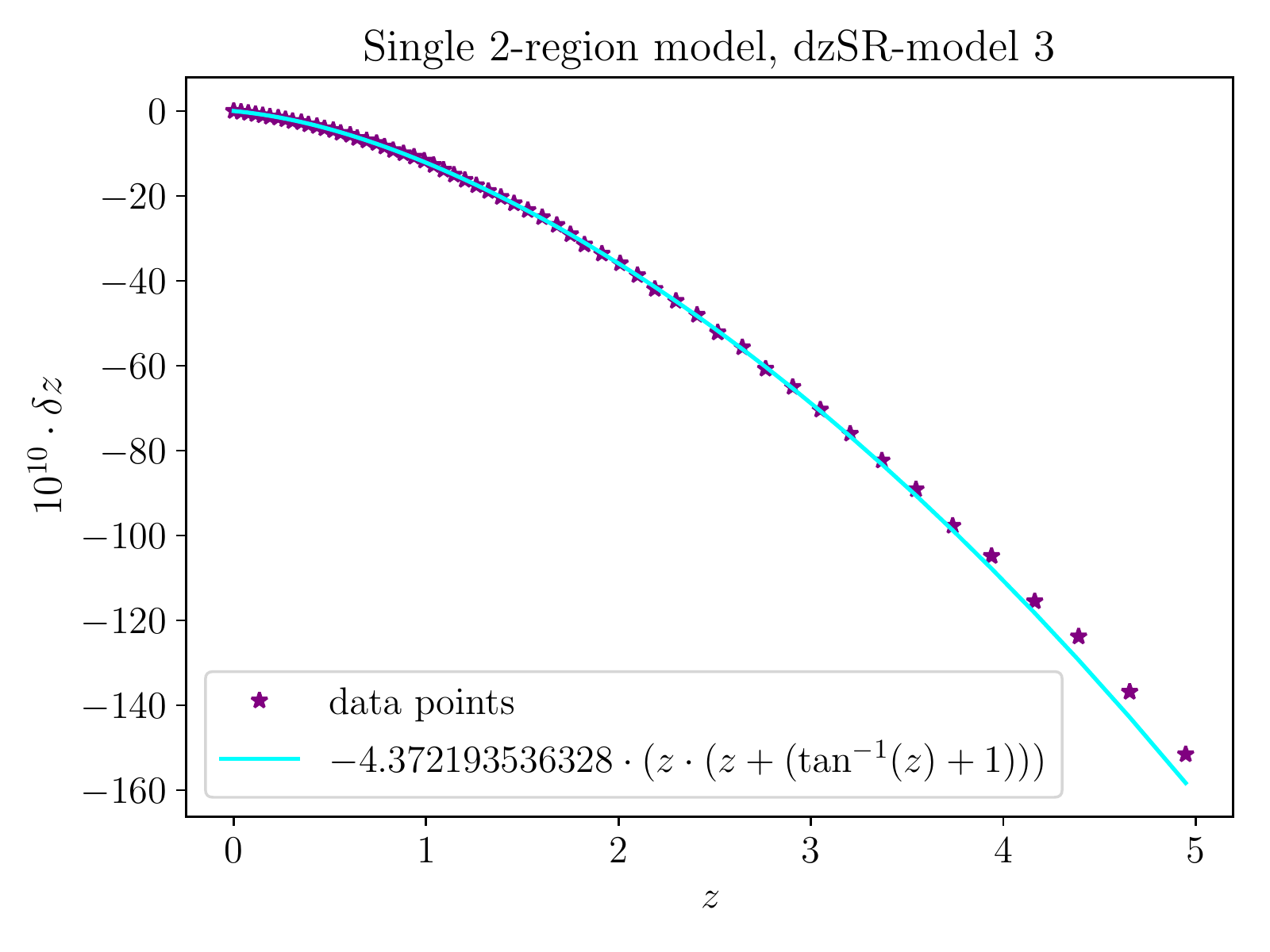}
	}
	\subfigure[]{
		\includegraphics[scale = 0.5]{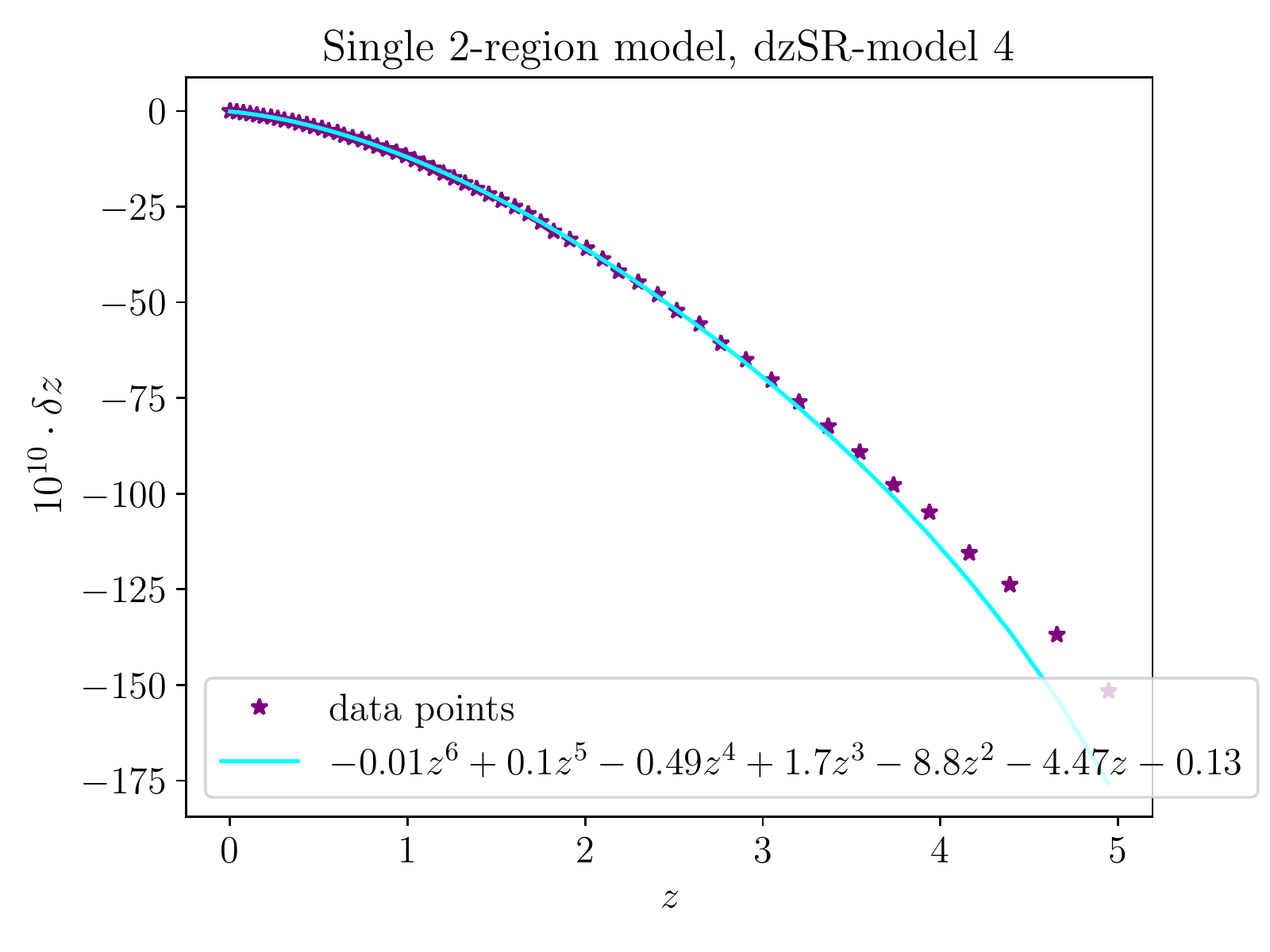}
	}
	\caption{Redshift drift data, $(z,\delta z)$, and a selection of symbolic expressions obtained with AI Feynman.}
	\label{fig:dz_2region}
\end{figure*}
\begin{figure}
\centering
	\includegraphics[scale = 0.5]{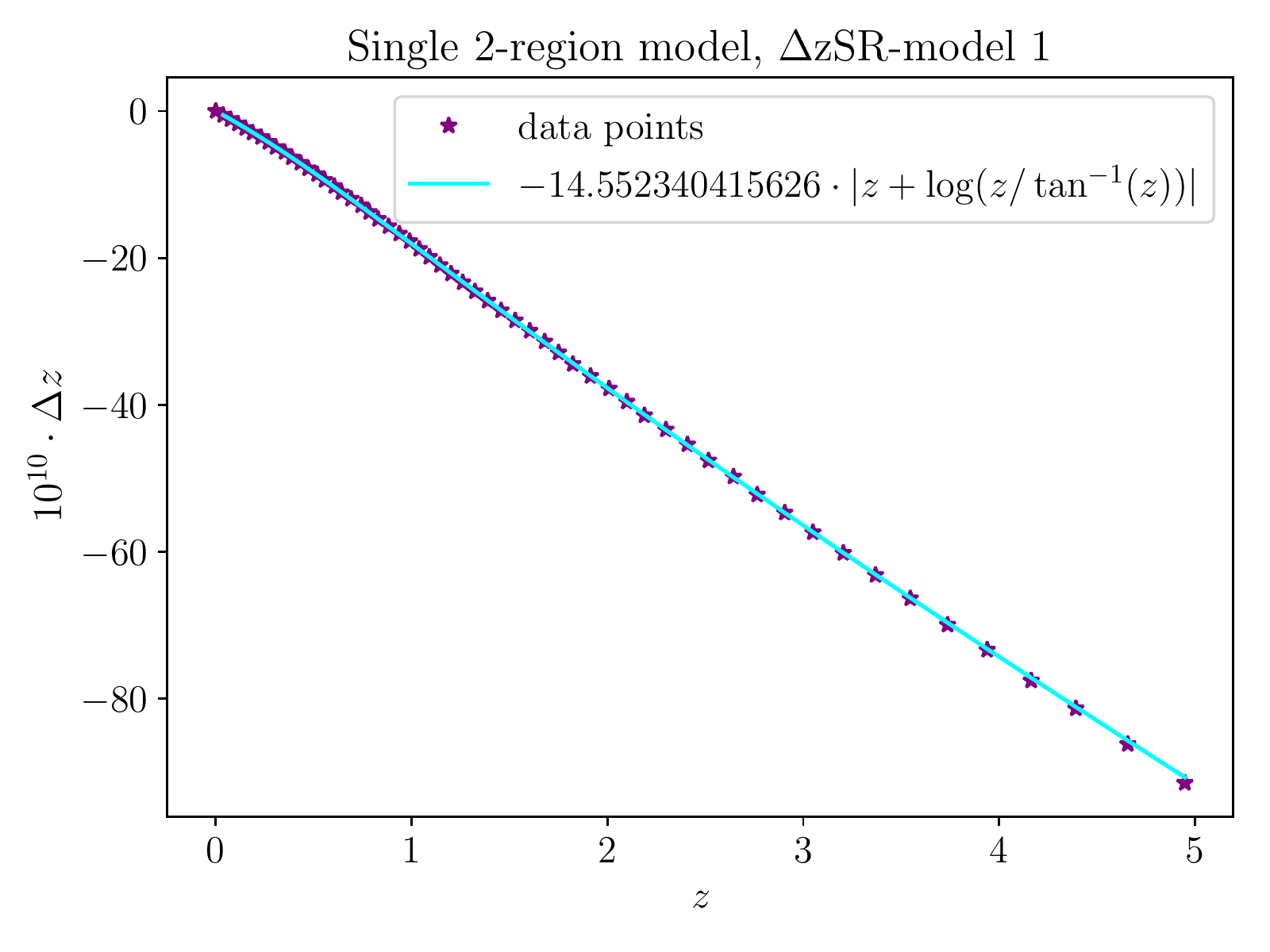}
\caption{Symbolic expressions for $(z,\Delta z)$ found by AI Feynman plotted together with data points.}
\label{fig:fit_Ddz}
\end{figure}
\begin{figure}
	\centering
	\subfigure[]{
		\includegraphics[scale = 0.5]{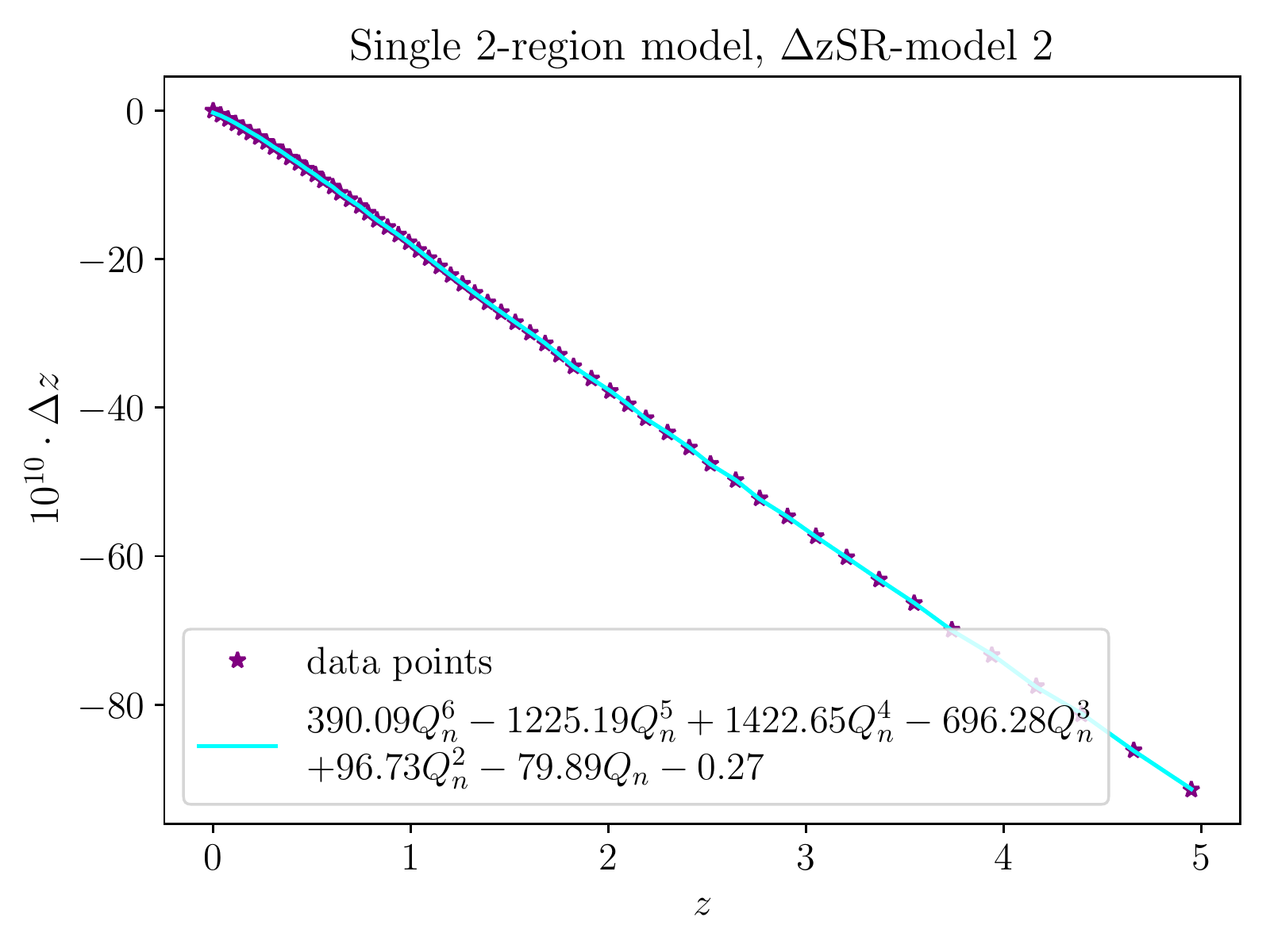}
	}\par
	\subfigure[]{
		\includegraphics[scale = 0.5]{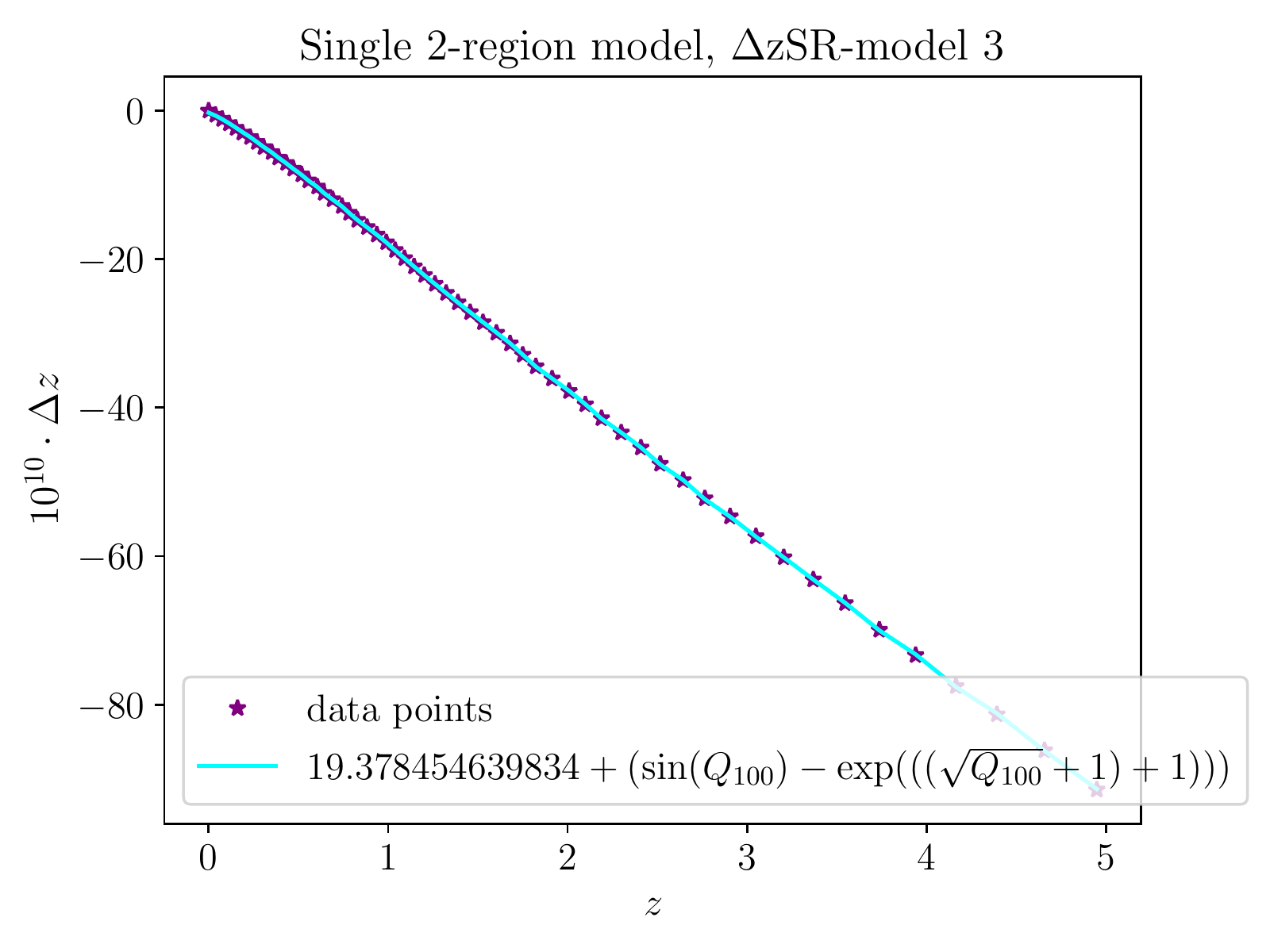}
	}
	\caption{Symbolic expression for $(Q_{D,x}, \Delta z)$, where $x = n,100$ indicates scaling of $Q$ either through a normalization or a simple scaling by a factor of 100.}
	\label{fig:Ddz_normalized}
\end{figure}
The results from presenting AI Feynman with redshift drift data from the single 2-region model are shown in figure \ref{fig:dz_2region}. Especially the expression marked as dzSR-model 1 is reasonably accurate -- indeed, its accuracy is roughly the same as the total expressions obtained after two rounds of using AI Feynman on the $Q$ data (figure \ref{fig:SecondFit}), with an accuracy of a few percent or below up to around $z = 4$.
\newline\newline
We could now move on to do a second round of using AI Feynman on the $(z,\delta z)$ data as with the kinematical backreaction, to try to obtain a fit with higher accuracy for the high-redshift part of the data. However, we will skip this and instead note that if the mean redshift drift were equal to the drift of the mean redshift, we would have the relation
\begin{align}\label{eq:dz_naive}
\left\langle \delta z\right\rangle  = \delta \left\langle z\right\rangle  = \delta t_0\left[ (1+z)H_{D_0} - H_D \right] ,
\end{align}
which is equivalent to the FLRW expression for the redshift drift. Remember that triangular brackets indicate taking the mean over several random lines of sight. Hence, $\left\langle \delta z\right\rangle $ is the mean redshift drift while $\delta \left\langle z\right\rangle $ is the drift of the mean redshift.
\newline\indent
As shown in \cite{another_look}, the mean redshift drift and the drift of the mean redshift are not identical in 2-region models so the first equality in the above expression does not hold. It is then interesting to see if there is another relationship between $\left\langle \delta z\right\rangle $ and $H_D$, $z$ and $Q$. This might seem a hopeless venture (or at least a venture requiring a lot of tuning of the algorithm) because $z$ and $H_D$ are not independent and because AI Feynman does not know that $\left\langle \delta z\right\rangle $ reduces to $\delta \left\langle z\right\rangle $ in the FLRW limit. Thus, even if $H_D$ were added as a feature, there is no reason to expect that the resulting expression would resemble equation \ref{eq:dz_naive}. However, since \emph{we} know that in the FLRW limit, $\left\langle \delta z\right\rangle  = \delta \left\langle z\right\rangle $, it would be desirable that the symbolic expressions obtained with AI Feynman reflect this. To achieve this, AI Feynman was presented to first data of the type $(z,\Delta z)$ and afterwards to data of the type $(Q, \Delta z)$, where $\Delta z:=\left\langle \delta z\right\rangle -\delta\left\langle z\right\rangle $. Then, $\left\langle \delta z\right\rangle $ can be written as the sum of $\delta \left\langle z\right\rangle $ and the symbolic expression found by AI Feynman. This turns out to be an easier task for AI Feynman in the sense that the algorithm is fast to find several symbolic expressions for $\Delta z$ in terms of $z$ with sub-percent accuracy on the main part of the interval. An example is shown in figure \ref{fig:fit_Ddz} which has an accuracy below 1\% except for approximately $z\leq 0.6$ where the inaccuracy increases to almost 2\%. However, the other data set, $(Q, \Delta z)$, is more interesting since the literature so far strongly suggests that $\Delta z$ will deviate significantly from zero only when there is significant backreaction \cite{another_look, Hellaby, dzLTB, in_progress} (or if the studied model clearly does not have hypersurfaces with statistical homogeneity and isotropy as in e.g. \cite{dzLTB_void, dzSzekeres1, dzSzekeres2, dzStephani, dz_Bianchi}). It turns out also to be much more difficult to obtain an accurate symbolic expression for this data with AI Feynman: Several runs with AI Feynman using different input choices (time limit for the brute force call, different symbolic expression files etc.) were all unsuccessful in finding accurate expressions. In an attempt to try to obtain more accurate expressions, the $Q$ values were normalized according to $Q\rightarrow Q_{\rm n}:= (Q-Q_{\rm min})/(Q_{\rm max}-Q_{\rm min})$. The most accurate expression obtained this way is shown in figure \ref{fig:Ddz_normalized}. The expression is accurate to 0.1\% level except for at very low redshifts where it reaches percent-level for $z\leq0.2$. However, as discussed earlier, using normalized data is inconvenient. AI Feynman was therefore also used on a data set where $Q$ was simply scaled by a factor of 100. In this case, it is again possible to obtain symbolic expressions with percent-level accuracy, presumably because this simple scaling roughly brings $Q$ to the same order of magnitude as the other features for a larger part of the feature intervals. An example of an accurate expression obtained by scaling $Q$ by a factor of 100 is shown in figure \ref{fig:Ddz_normalized}. The expression has a sub-percent accuracy for $0.2\leq z\leq2.5$ but outside this interval it slowly increases to reach 10\%-order for the smallest and largest values of the redshift in the studied interval.
\newline\indent
With these expressions, we can write $\left\langle \delta z\right\rangle $ in the form
\begin{align}
 	\left\langle \delta z\right\rangle  = \delta t_0\left[ (1+z)H_{D_0} - H_D \right] + F(Q),
\end{align}
where $F(Q)$ is the symbolic expression found for $\Delta z$ in terms of $Q$.
\newline\newline
Before closing this section, a note on the phenomenological nature of the symbolic expressions is appropriate: The expressions obtained with AI Feynman cannot {\em a priori} be expected to represent physically justifiable expressions but are instead phenomenological models that cannot generally be extrapolated to outside the feature intervals used for obtaining the expressions. This is true regardless of the form of the symbolic expressions but is perhaps emphasized well for polynomial expressions since especially polynomials of high degree are well-known for being able to (over-)fit data on smaller intervals very accurately. This is worth remembering when regarding the polynomial expression in figure \ref{fig:Ddz_normalized}; the AI Feynman algorithm was instructed to use $6$ as the maximum polynomial power when it obtained this expression. This is also exactly the degree of the obtained polynomial. The possibility for obtaining physically justifiable expressions through symbolic regression is discussed in section \ref{sec:conclusion}. 
\begin{figure}
	\centering
		\includegraphics[scale = 0.5]{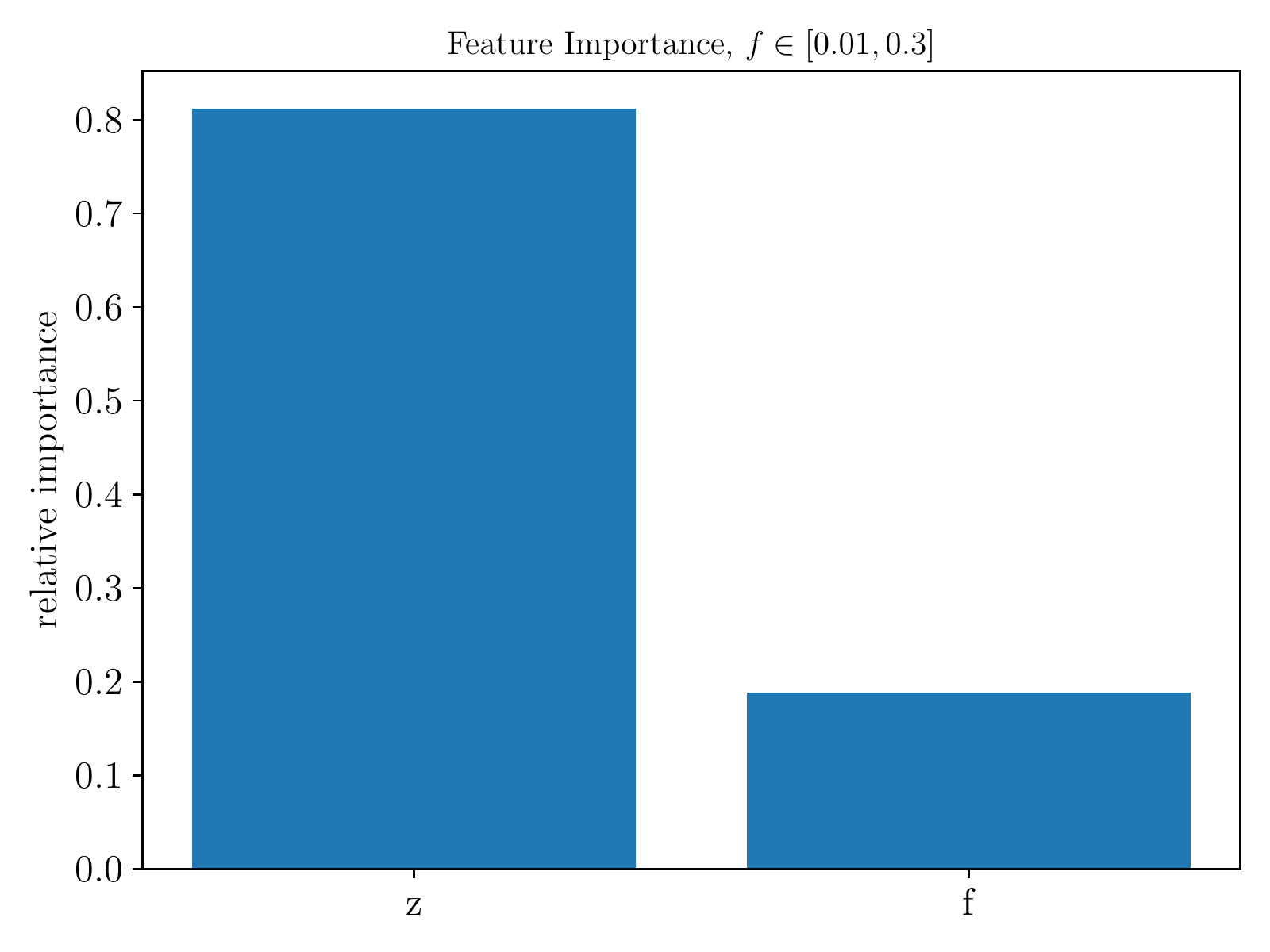}
	\caption{Relative importance of the independent features in the backreaction data.}
	\label{fig:feature_Q_final}
\end{figure}
\label{subsec:feature}
\begin{figure}
	\centering
	\includegraphics[scale = 0.5]{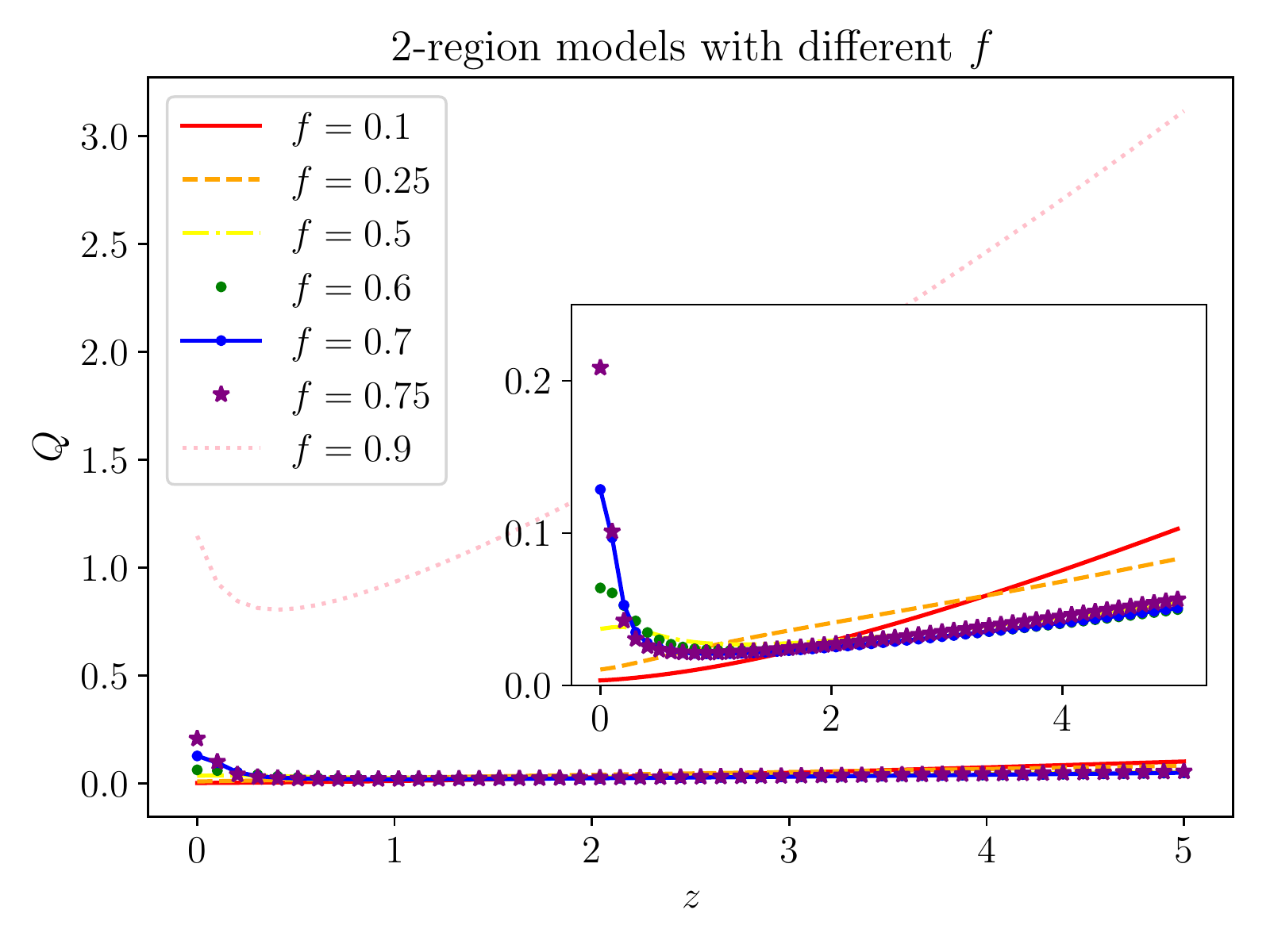}
	\caption{Kinematical backreaction as a function of the redshift for different values of $f$. A close-up is included to better show the different $Q(z)$ for $f\neq 0.9$.}
	\label{fig:differen_f}
\end{figure}
\begin{figure*}
	\centering
	\subfigure[]{
		\includegraphics[scale = 0.5]{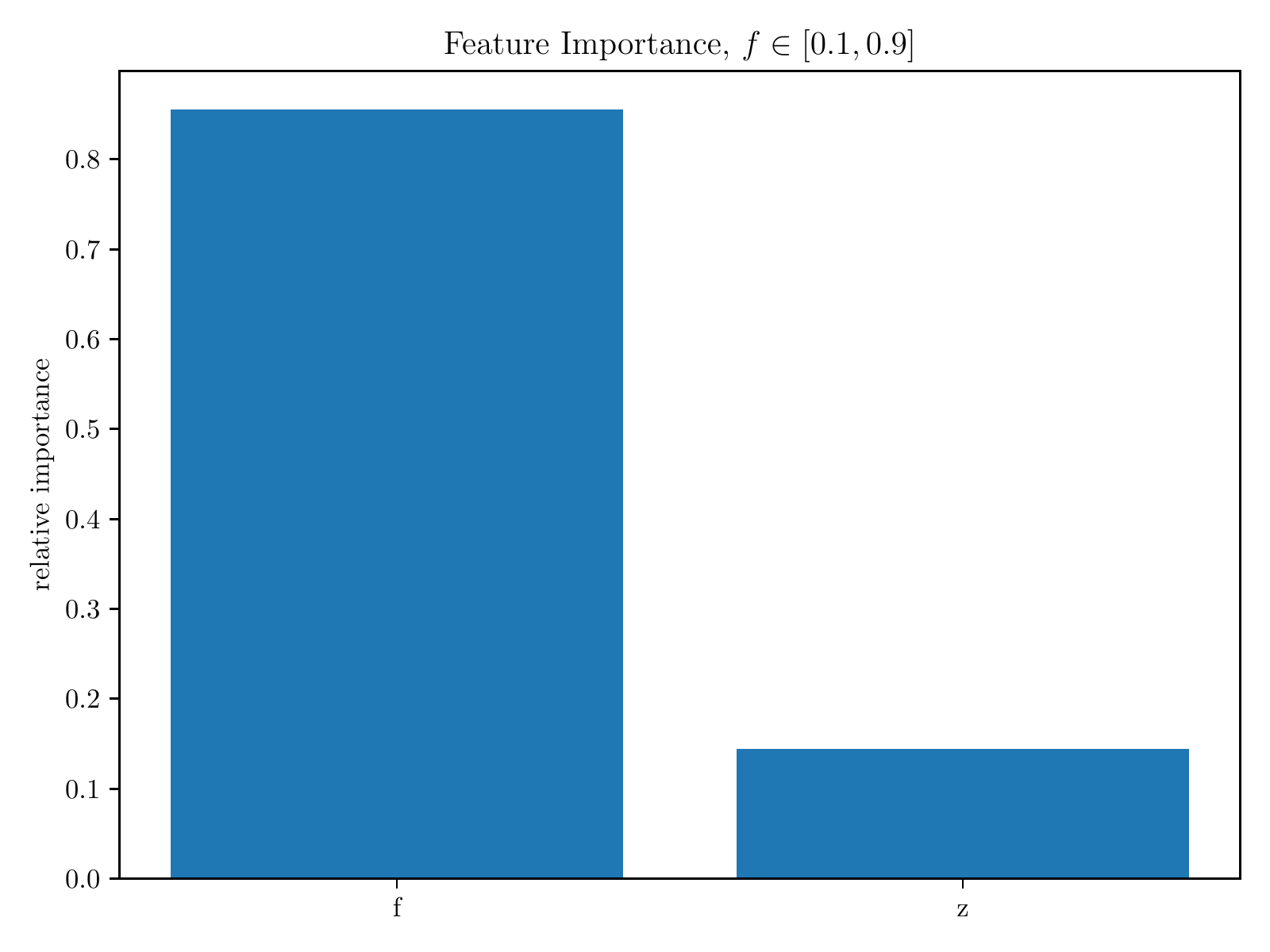}
	}
	\subfigure[]{
		\includegraphics[scale = 0.5]{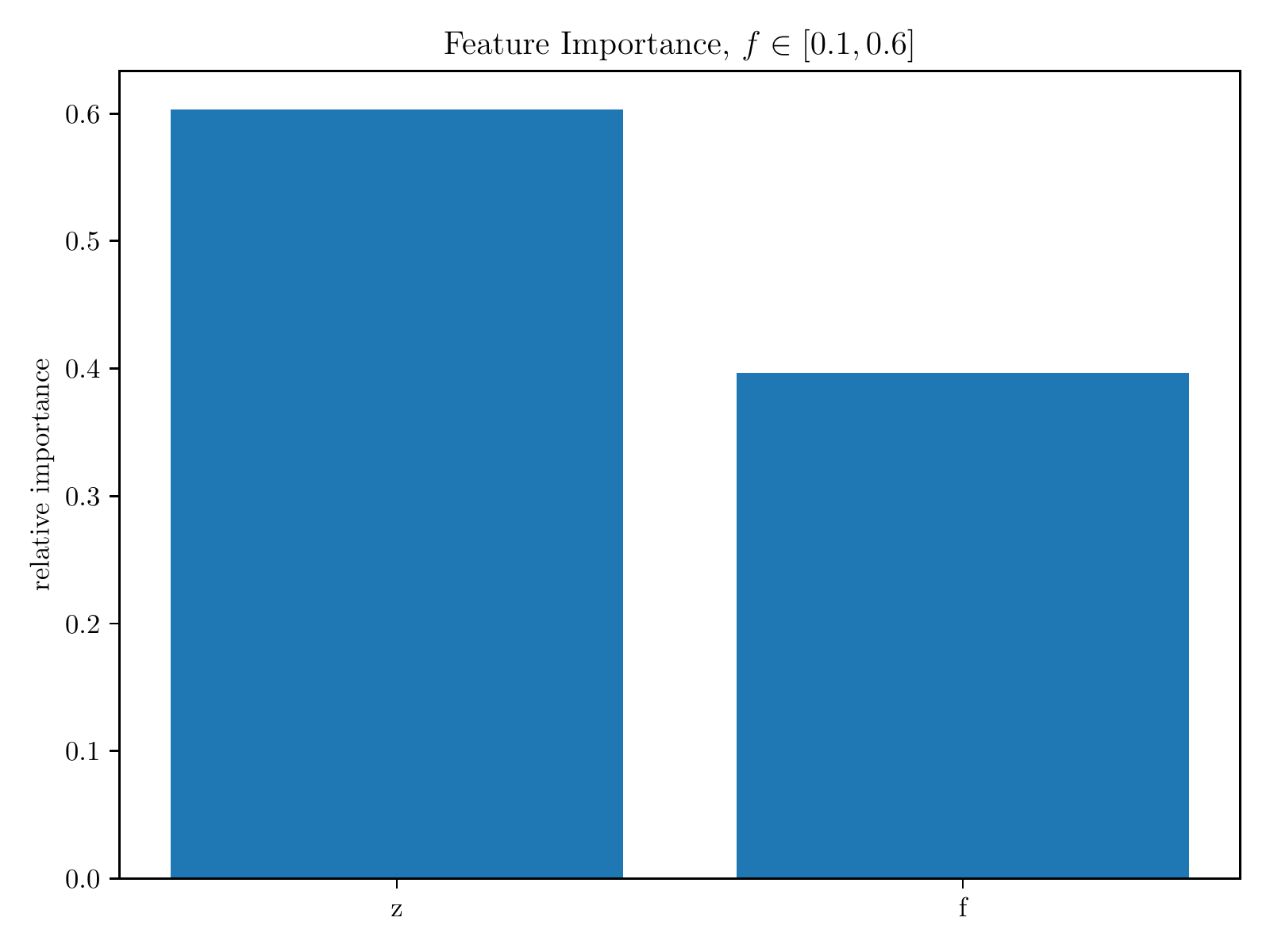}
	}
	\caption{Relative importance of the independent features in the backreaction data. Note that features are arranged such that the most important feature is always shown furthest to the left and so forth.}
	\label{fig:feature_Q}
\end{figure*}

\subsection{Multiple 2-region models: Feature importance}
\begin{figure}
	\centering
	\includegraphics[scale = 0.5]{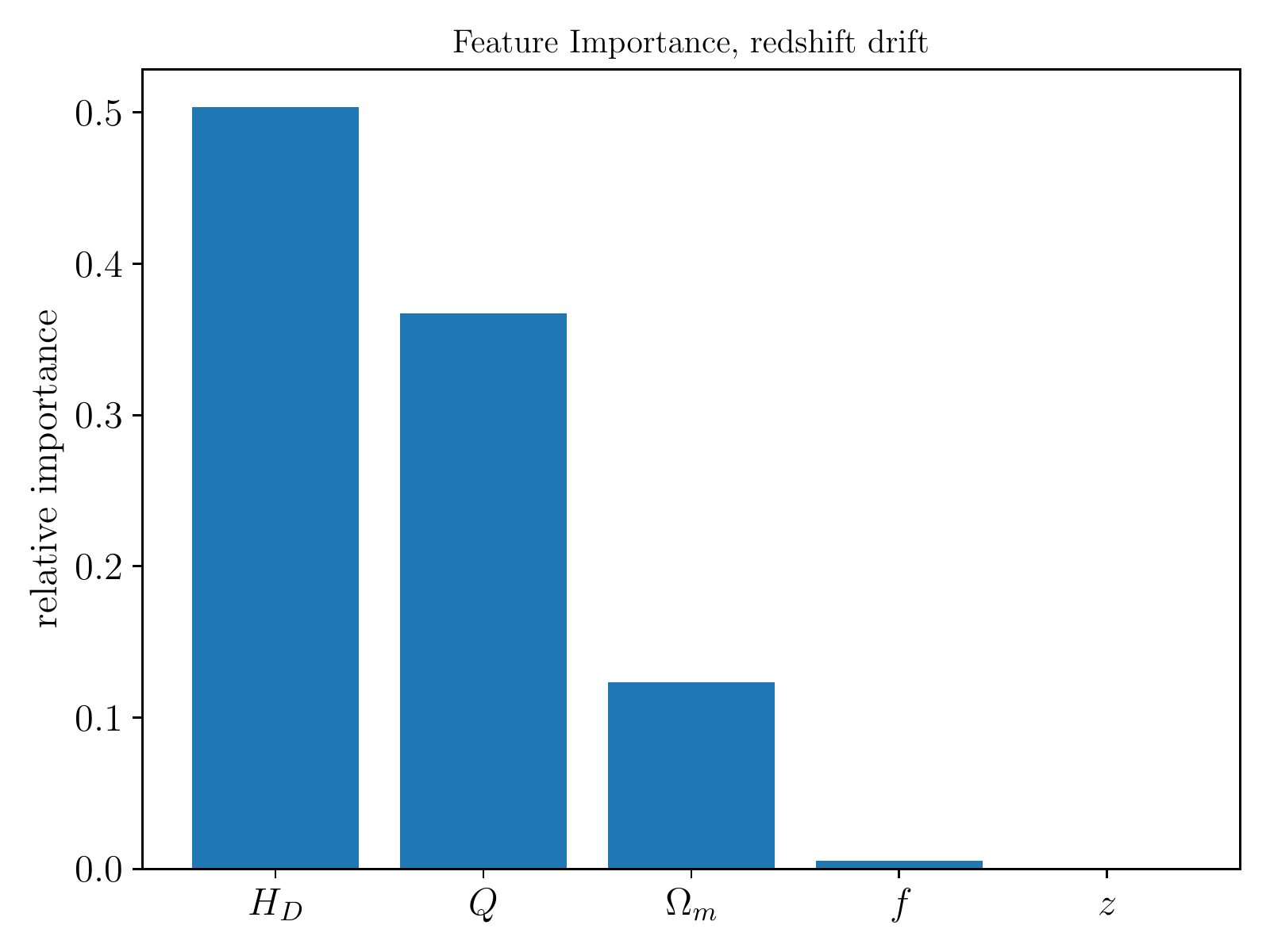}
	\caption{Relative importance of the different features in the redshift drift data.}
	\label{fig:feature_importance_dz}
\end{figure}
\begin{figure}
	\centering
	\includegraphics[scale = 0.5]{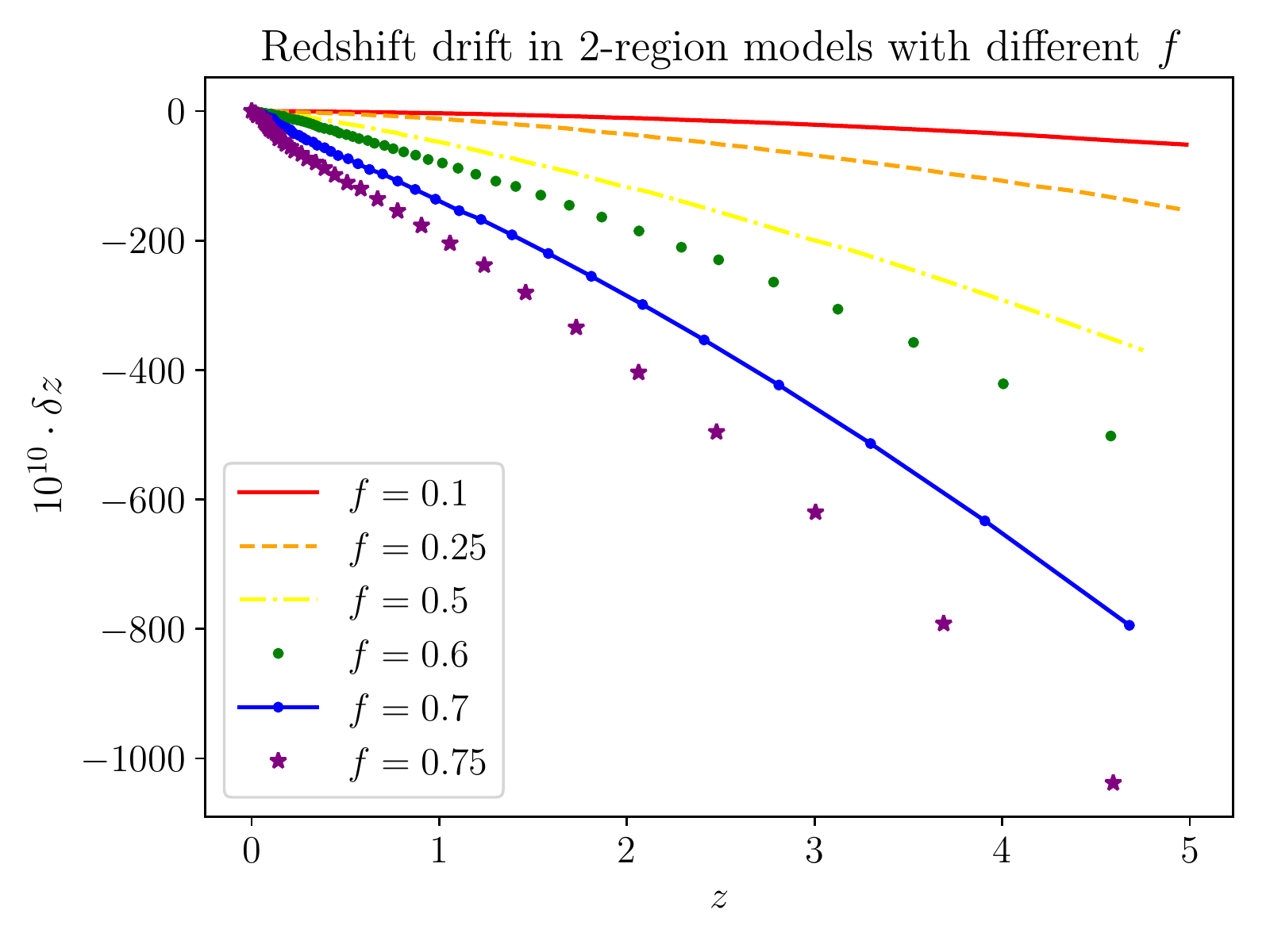}
	\caption{Redshift drift as a function of the redshift for different values of $f$.}
	\label{fig:dz_differen_f}
\end{figure}
\begin{figure}
	\centering
	\includegraphics[scale = 0.5]{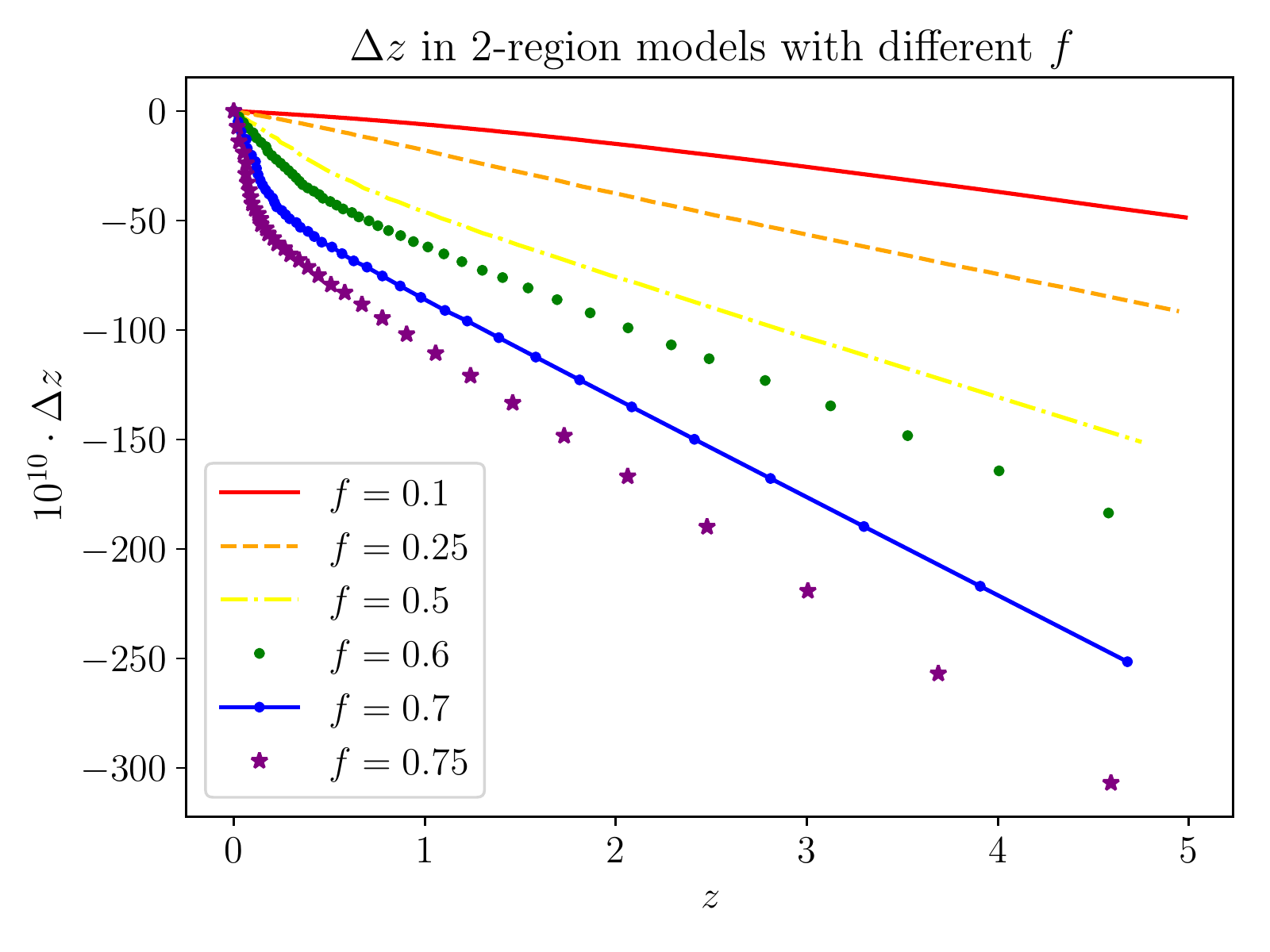}
	\caption{The deviation, $\Delta z$, as a function of the redshift for different values of $f$.}
	\label{fig:ddz_differen_f}
\end{figure}
Doing symbolic regression on the kinematical backreaction and redshift drift of a single 2-region model can be useful as an initial exercise since the results are informative on what can be expected regarding a successful regression analysis on data from multiple models. However, this latter regression analysis is the real goal: Understanding to what extent $Q$, $R_D$ and $\left\langle \delta z\right\rangle $ can be parameterized in terms of spatially averaged quantities with general expressions valid for a significant parameter/feature region. Before presenting the results obtained by presenting AI Feynman with data from multiple 2-region models, we will in this subsection study the relative importance of the individual features for the values of the target variables $Q$ and $\left\langle \delta z\right\rangle $ in multiple models (we will defer considering $R_D$ until next subsection). Feature importance is a standard machine learning tool that can be used to learn about which features of a data set the target depends (the most) on and hence guide the use of which feature to focus e.g. a regression task on (see e.g. \cite{ML_bog}). We could therefore do a feature importance study where different average quantities such as $H_D$, $\rho_D$, $z$, $a_D$ etc. were all added as features for the targets $Q$ and $\delta z$. The feature importance study would then show us which of these features contain the most information about the targets and hence are the most prudent to include as features for the target when doing the later regression. However, we are here working with a toy-model where the differences between different 2-region models are actually contained in a single feature, namely $f$. In addition, the goal here is specifically to parameterize $Q$ in terms of the volume averaged scale factor or corresponding redshift. Therefore, for $Q$, the only features that will be considered are $f$ and $z$. It is still instructive to compute the relative importance of these two features since the features have very different physical meanings with one feature being a model parameter while the other is an observable quantity. In addition, we learned in the previous section that accurate expressions can be obtained for $Q$ and $\delta z$ for individual 2-region models so the difficulties we can expect to encounter when moving to multiple 2-region models may depend on whether $z$ or $f$ is the most important feature.
\newline\indent
The feature importances are computed using scikit-learn \cite{scikit}. The results shown in this subsection are based on feature importances extracted from a random forest regressor with 100 estimators (trees). For the random forest regressors, feature importances are estimated through combinations of the standard deviation and mean of the accumulated impurity increase of each tree. See e.g. \cite{ML_bog} for an introduction to random forests, decision trees and feature importance\footnote{The reader may also find it instructive to look directly at the scikit-learn page https://scikit-learn.org/stable/auto\_examples/ensemble/plot\_forest\_importances.html.}.
\newline\newline
The relative importance of $f$ and $z$ for the value of $Q$ are shown in figure \ref{fig:feature_Q_final}. For the computation, the interval of $f$ was initially set to $f\in[0.01,0.3]$ while $z\in[0,5]$. The interval for $f$ was chosen based on noting that for $f\approx 0.18$, the kinematical backreaction is at the same order as the cosmological constant in the standard $\Lambda$CDM model and for $f\approx 0.225$, the model leads to a redshift-distance relation very similar to the redshift-distance relation of the standard model (the latter is shown in figure 2 of \cite{dz_with_Steen}). Thus, by searching an interval with $f\in[0.01,0.3]$ we are considering a large variation in the kinematical backreaction, centered roughly in an area that can under some circumstances mimic what is observed in the real universe regarding dark energy. It is nonetheless interesting to briefly consider models with larger values of $f$ in order to assess how (dis-)similar the graphs for $Q(z)$ look for different values of $f$; this (dis-)similarity gives an indication of the complexity we should expect regarding a symbolic expression that can cover large feature intervals. For this reason, $Q(z)$ has been plotted for a range of $f$ up to $f = 0.9$. This is shown in figure \ref{fig:differen_f}. It is seen that large values of $f$ lead to quite extreme behavior of $Q$. This would presumably be difficult to capture in a symbolic expression that simultaneously accurately describes the backreaction in models with smaller values of $f$. Indeed, by looking at figure \ref{fig:differen_f} it is tempting to expect that $f$ may become a more important feature for $Q$ than $z$ is if a large interval of $f$ is considered. This is confirmed in figure \ref{fig:feature_Q} where the relative importance of $z$ and $f$ are shown for larger ranges of $f$. When $f\in[0.1,0.9]$\footnote{Note that this large interval of $f$ is used merely for illustrating the significance of changing the interval of $f$ for the feature importance study. A value of $f = 0.9$ corresponds to a universe where the main volume fraction of the Universe is made up of overdense regions even at present time, which is clearly not realistic.}, $f$ is by far the most important feature, i.e. the value of $f$ is more important than $z$ for the value of $Q$. When the interval is narrowed to $f\in[0.1,0.6]$, $z$ has become the most important feature. Note that feature importance is throughout depicted with the most significant feature furthest to the left and so forth. Because of this, $z$ and $f$ change places in the two diagrams shown in figure \ref{fig:feature_Q}.
\newline\newline
We will now turn to look at the redshift drift. As for the single 2-region model we will look at the dependence on the redshift and $f$, but the relative importance of $Q$, $\Omega_m:=8\pi G\rho_D/(3H_D^2)$ and $H_D$ will also be briefly discussed. The relative importance of all these features is shown in figure \ref{fig:feature_importance_dz}. In addition, the redshift drift for a variety of values of $f$ is shown in figure \ref{fig:dz_differen_f}.
\newline\indent
The relative importances depicted in figure \ref{fig:feature_importance_dz} indicate that $H_D$ by far is the most important feature for the value of the redshift drift -- and that the redshift is not really important at all. This is perhaps not too surprising, when remembering equation \ref{eq:dz_naive} and when looking at figure \ref{fig:dz_differen_f} where it is seen that for a given value of $f$, the redshift drift is quite flat along the $z$-axis compared to the change in the redshift drift between different models. Additionally, it is here important to remember that $H_D$, $\Omega_m$ and $Q$ all themselves depend on $z$ as well as on $f$. Therefore the relative importances in figure \ref{fig:feature_importance_dz} should not be considered without regarding the physical setup. What we are interested in physically is an expression for the mean redshift drift in terms of $z$ which is valid for the entire considered range of 2-region models. Therefore we should, as with $Q$, consider the redshift drift as a function of $z,f$. However, as discussed earlier, it is more desirable to obtain a symbolic expression of $\Delta z:=\left\langle \delta z\right\rangle -\delta\left\langle z\right\rangle $ in terms of $Q(z)$. This quantity is depicted in figure \ref{fig:ddz_differen_f} for a range of $f$ values. Based on the experience gained from considering a single 2-region model, $\Delta z$ will be considered with $z, Q_{100}$ and  $f$ as features in the following. It is nonetheless worth keeping in mind, e.g. for future development, that the feature importance study indicates that we can express much of the variability in the $\Delta z$ data by including $H_D$ and $\Omega_m$ as features.
\begin{figure*}
	\centering
	\subfigure[]{
		\includegraphics[scale = 0.35]{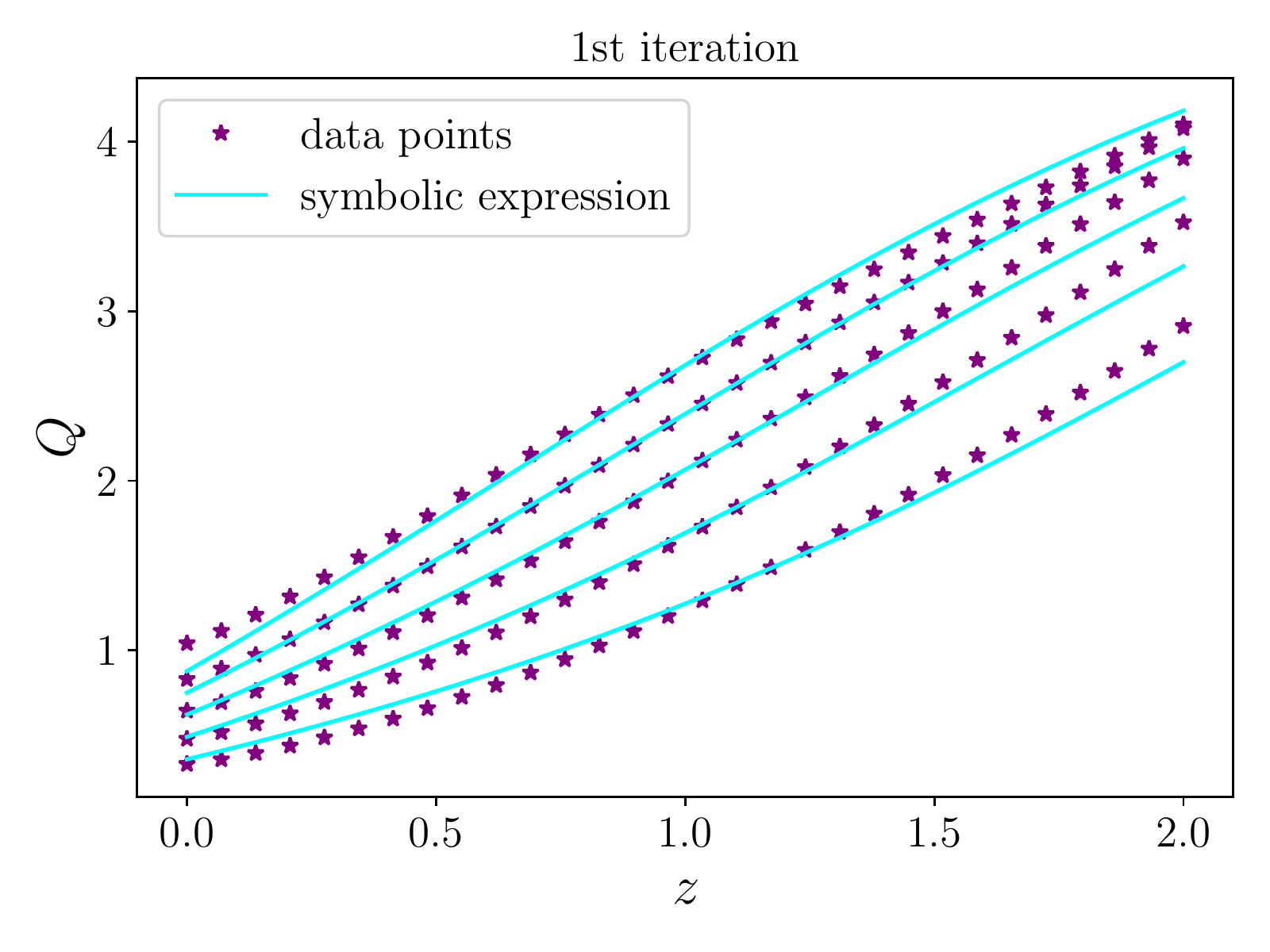}
	}
	\subfigure[]{
		\includegraphics[scale = 0.35]{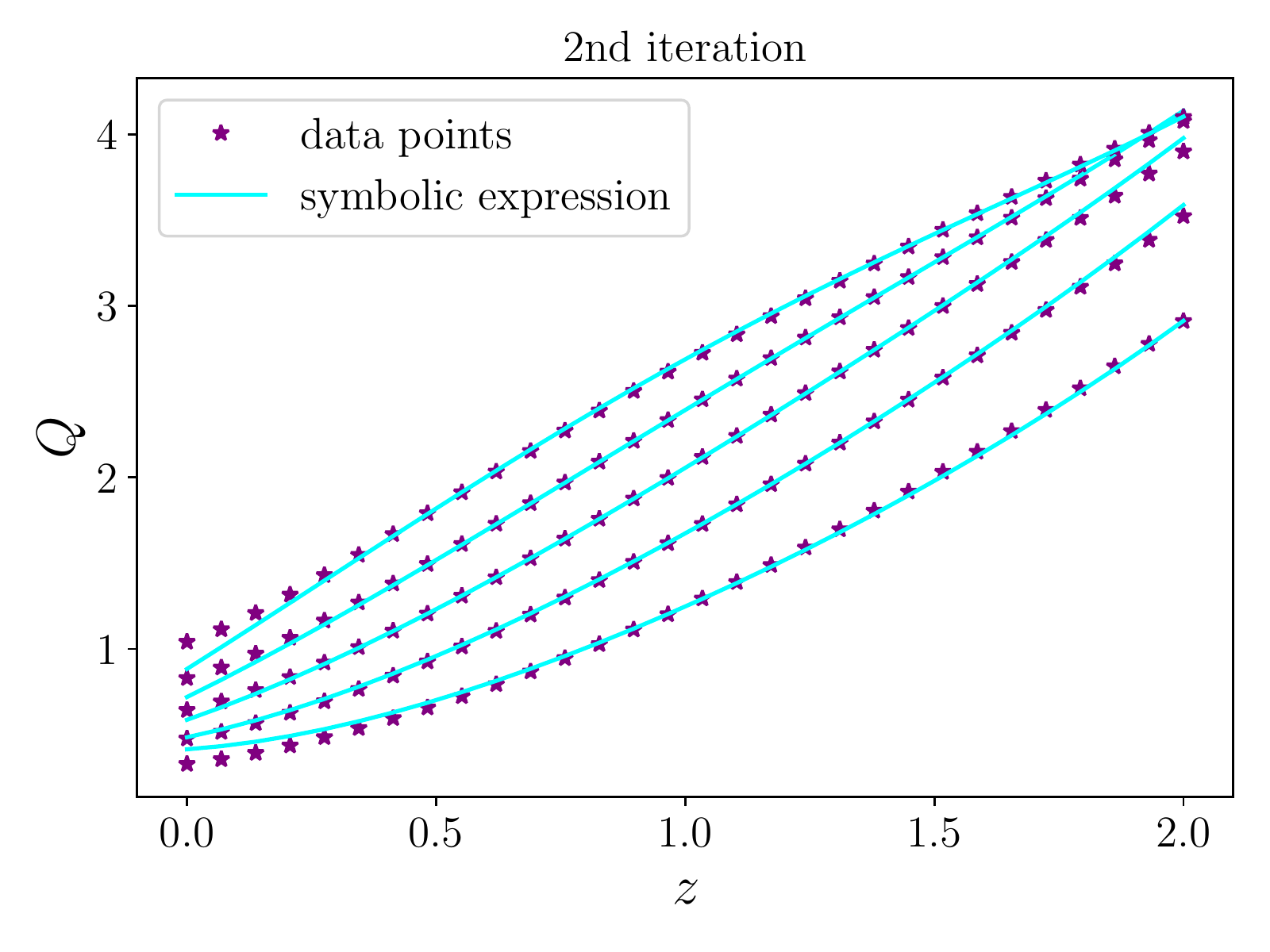}
	}
	\subfigure[]{
		\includegraphics[scale = 0.35]{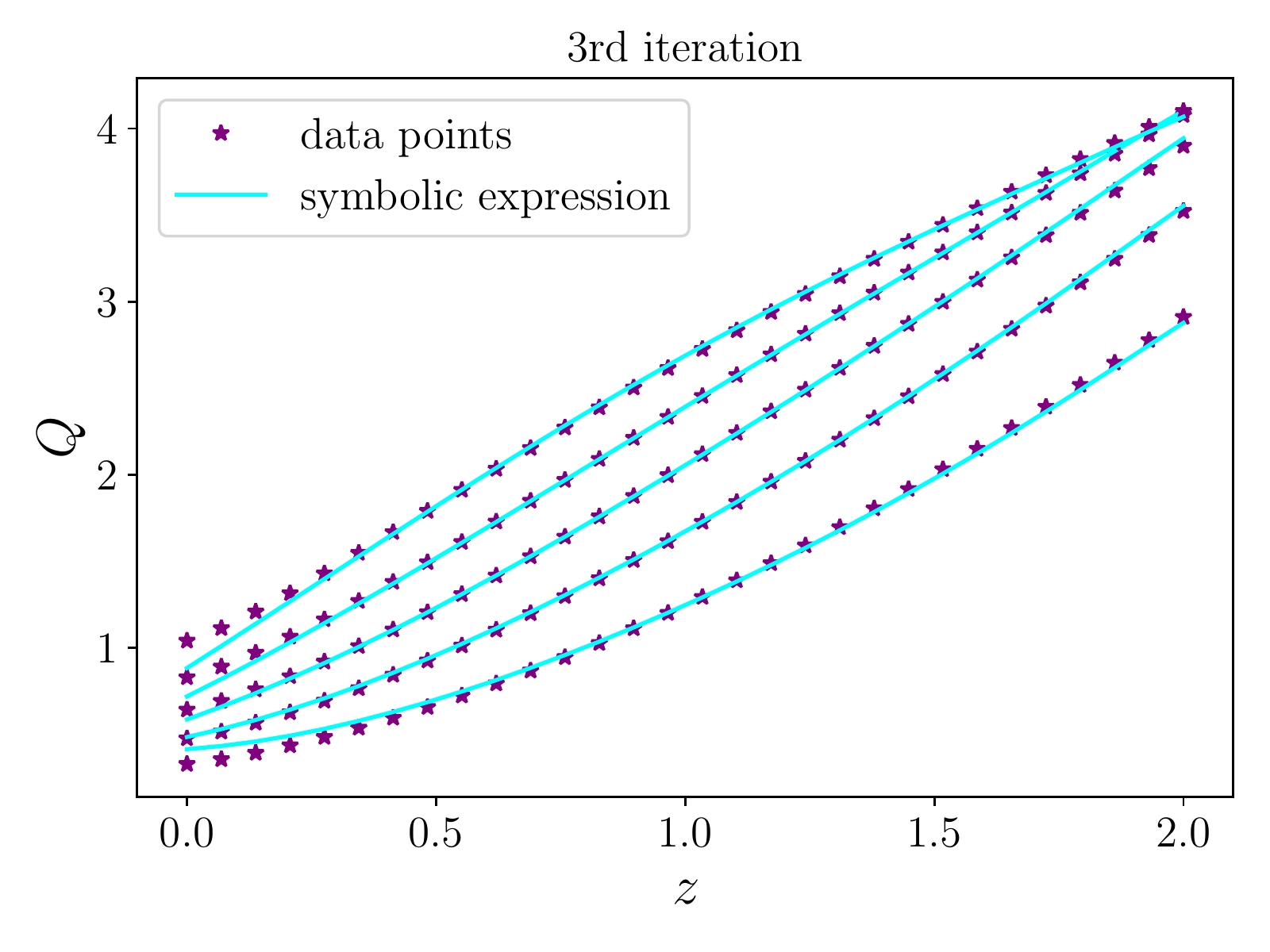}
	}\par
	\subfigure[]{
		\includegraphics[scale = 0.35]{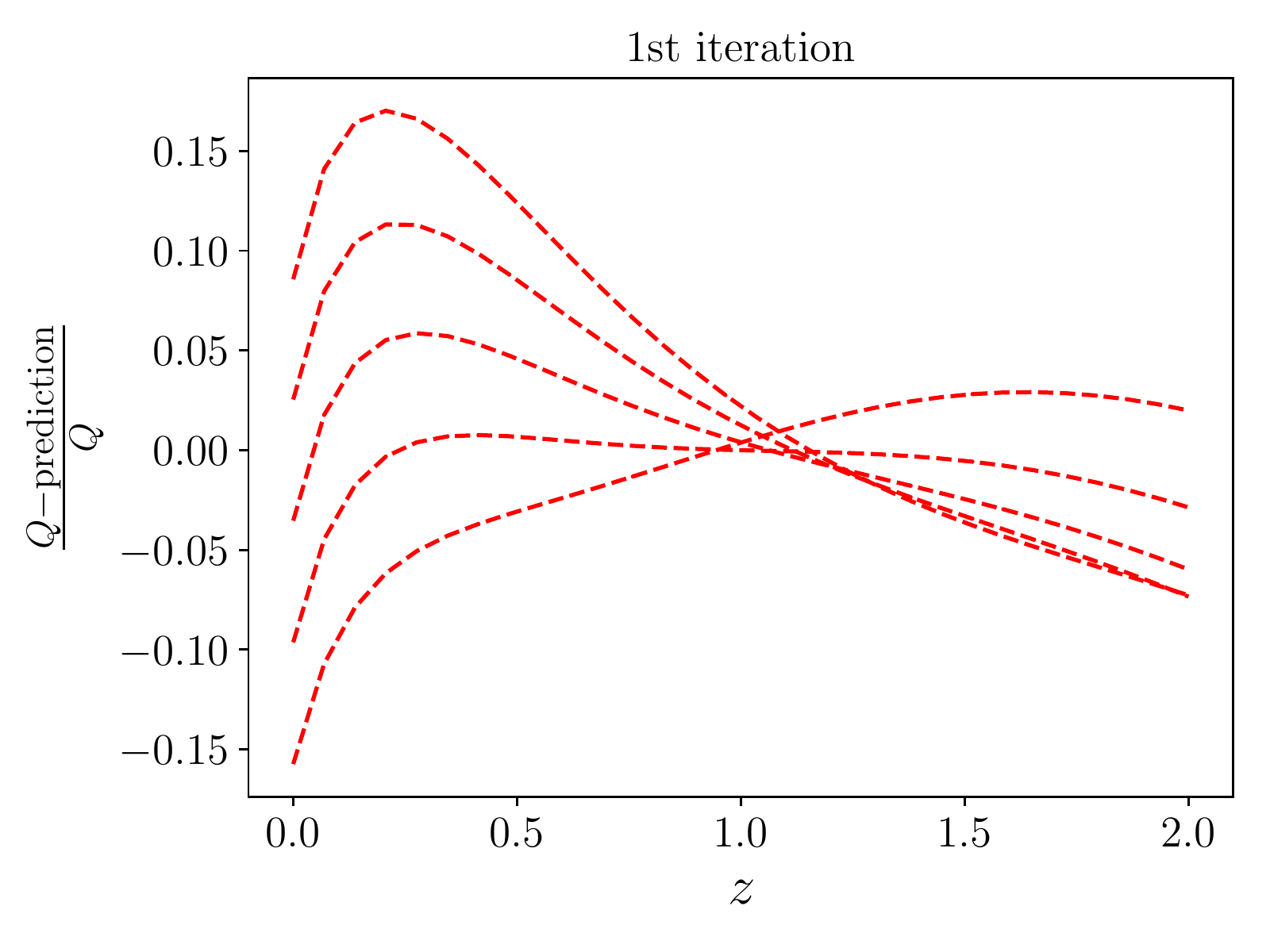}
	}
	\subfigure[]{
	\includegraphics[scale = 0.35]{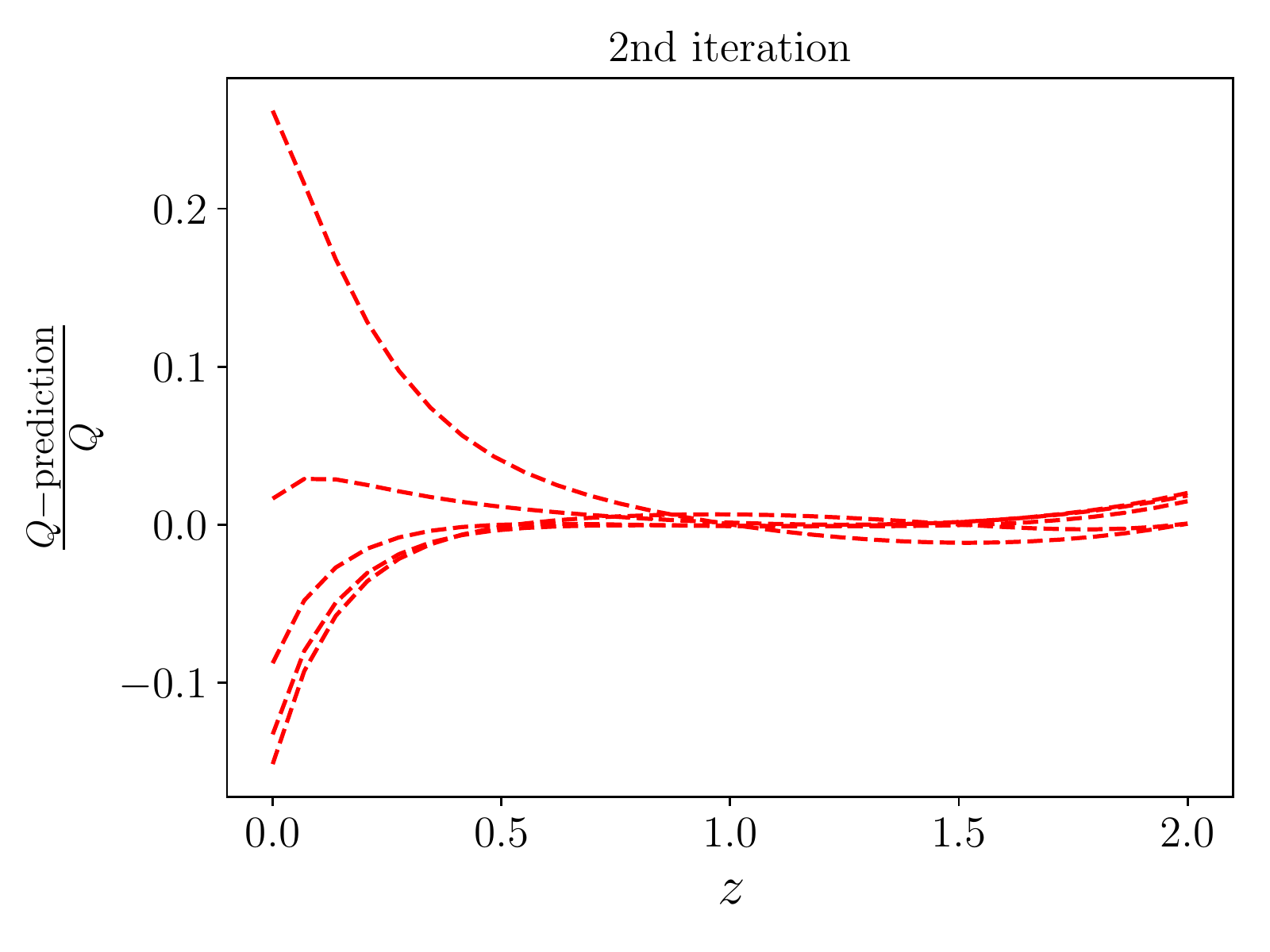}
	}
	\subfigure[]{
	\includegraphics[scale = 0.35]{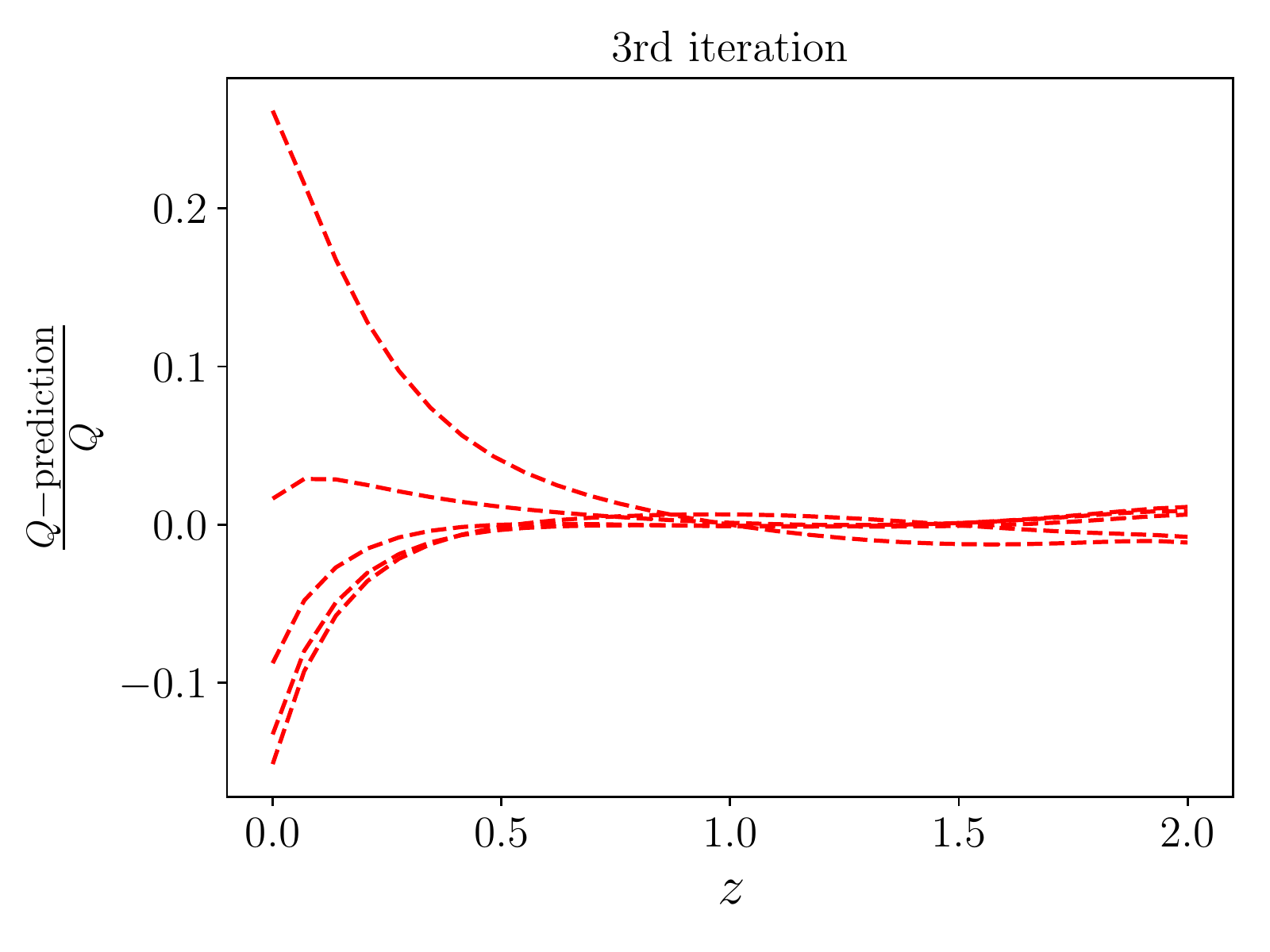}
	}
	\caption{Kinematical backreaction data $(z,f,Q)$ together with the most accurate expressions obtained with AI Feynman. The figure entitled ``1st iteration'' shows the data together with $f_1$, the figure entitled ``2nd'' iteration shows data together with $f_1+f_2$ etc. Figures are also included showing the relative deviation between data points and values according to the symbolic expressions, with the y-axis labeled as $\frac{Q-\rm prediction}{Q}$, where the prediction is the value obtained by evaluating the symbolic expression obtained from AI Feynman.}
	\label{fig:Qmult}
\end{figure*}

\begin{figure*}
	\centering
	\subfigure[]{
		\includegraphics[scale = 0.5]{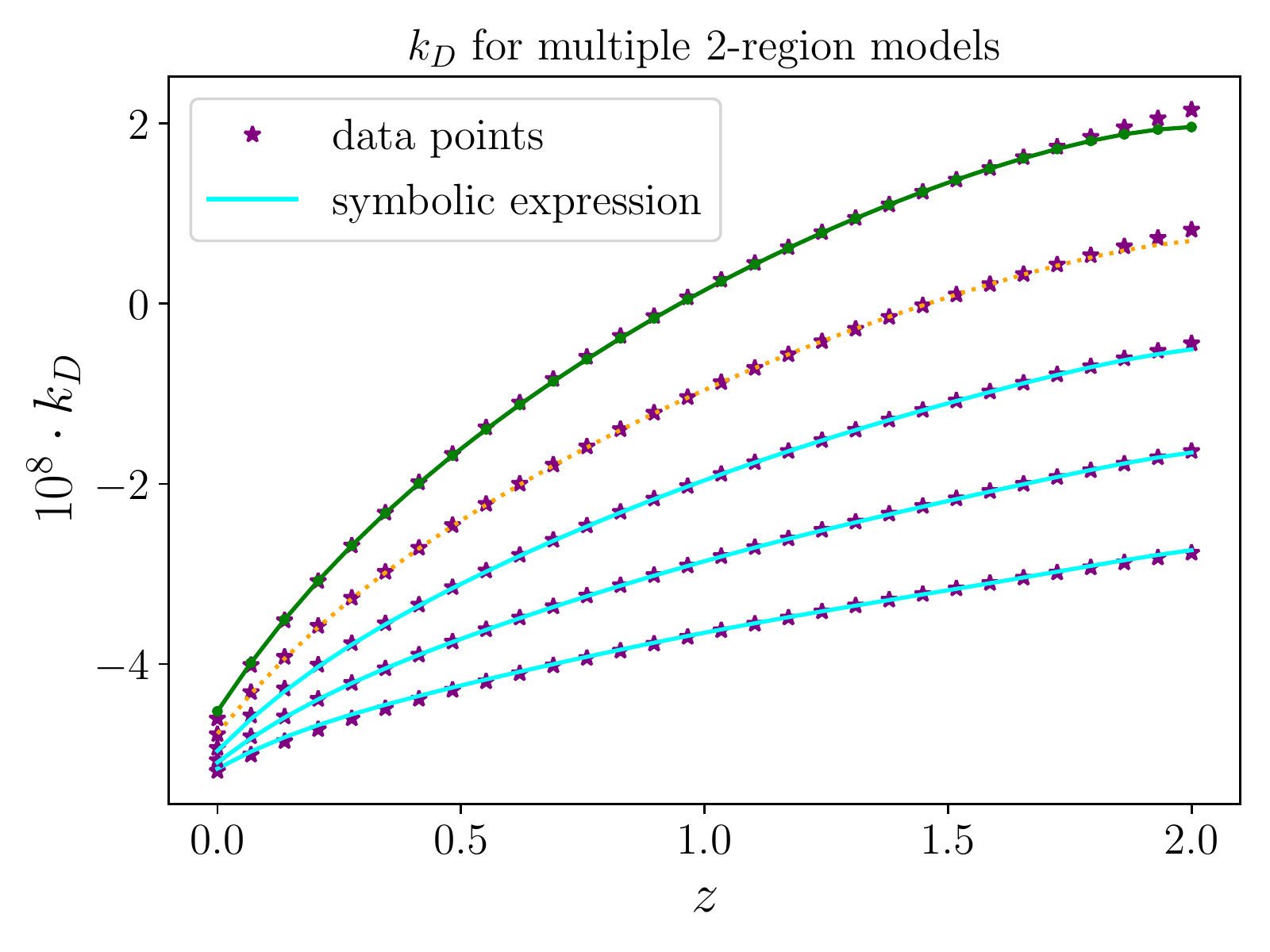}
	}
	\subfigure[]{
		\includegraphics[scale = 0.5]{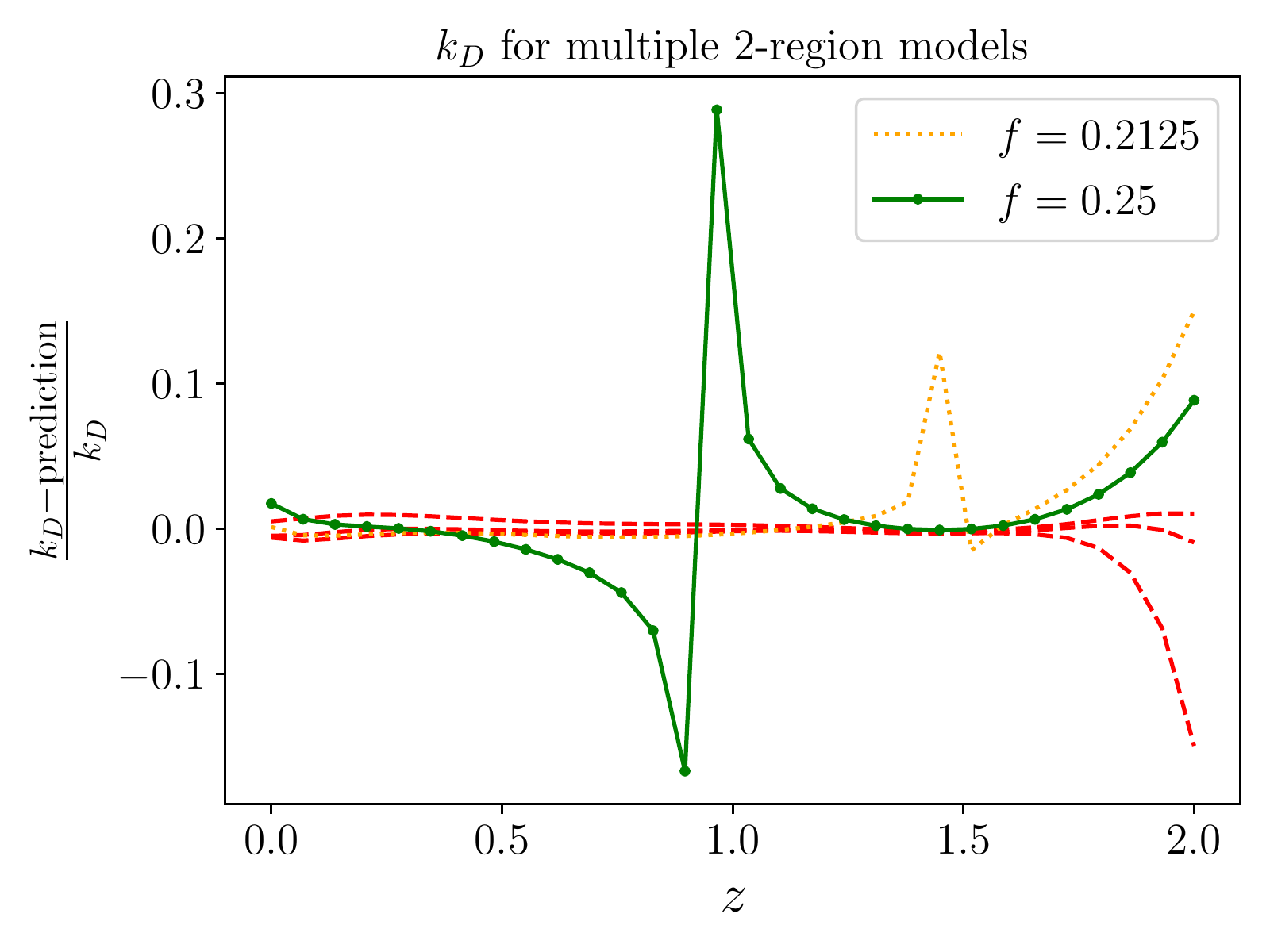}
	}	
	\caption{Curvature data $(z,f,10^{8}\cdot k_D)$ together with the most accurate expressions obtained with AI Feynman. Two lines are highlighted in the figures because they exhibit ``spikes'' in the relative precision. By comparing the two figures it is seen that these spikes come from division by zero because $k_D$ for these two models crosses zero i.e. the curvature changes sign in the models represented by those lines.}
	\label{fig:kmult_zeq2}
\end{figure*}

\begin{figure*}
	\centering
	\subfigure[]{
		\includegraphics[scale = 0.5]{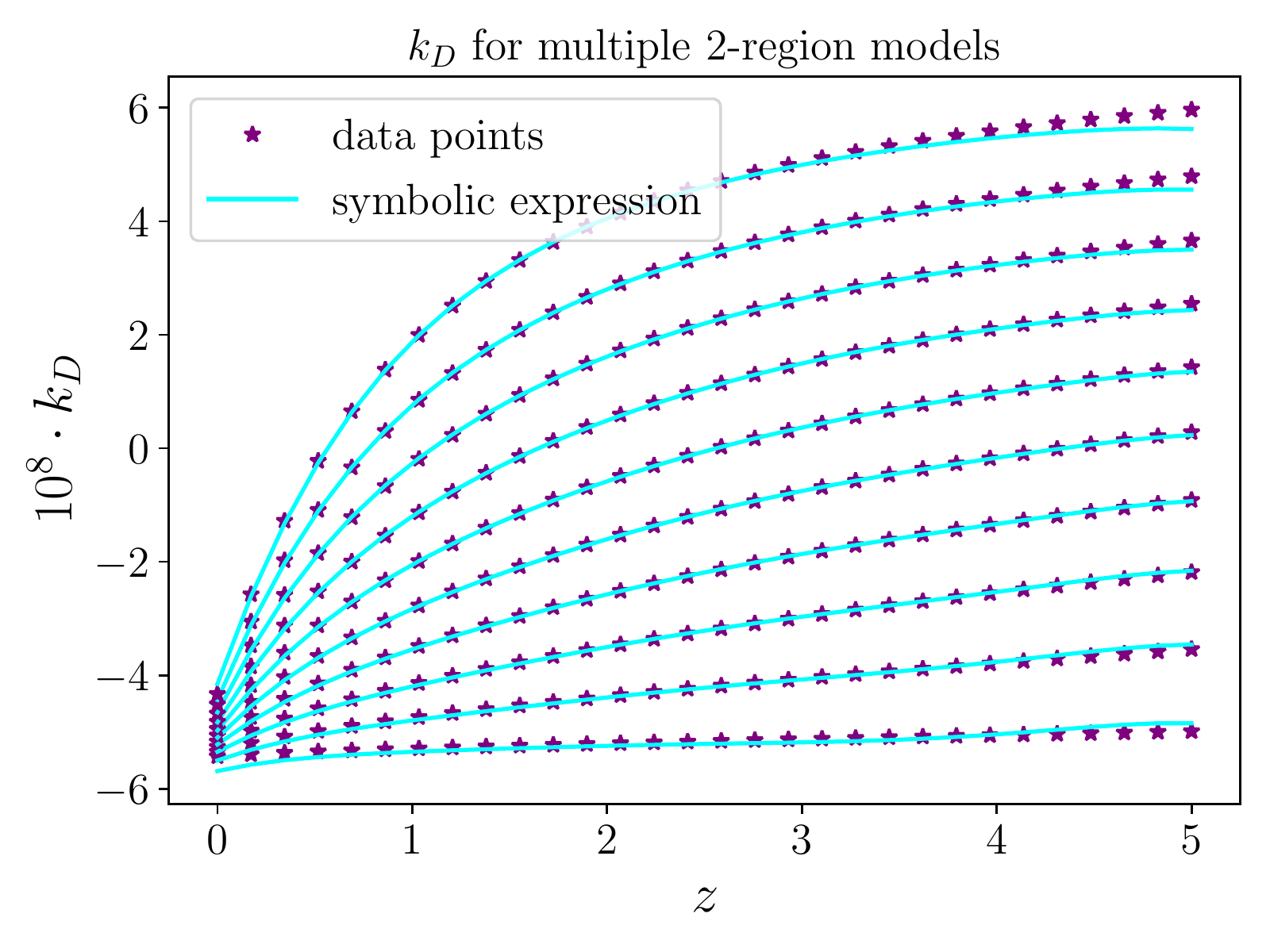}
	}
	\subfigure[]{
		\includegraphics[scale = 0.5]{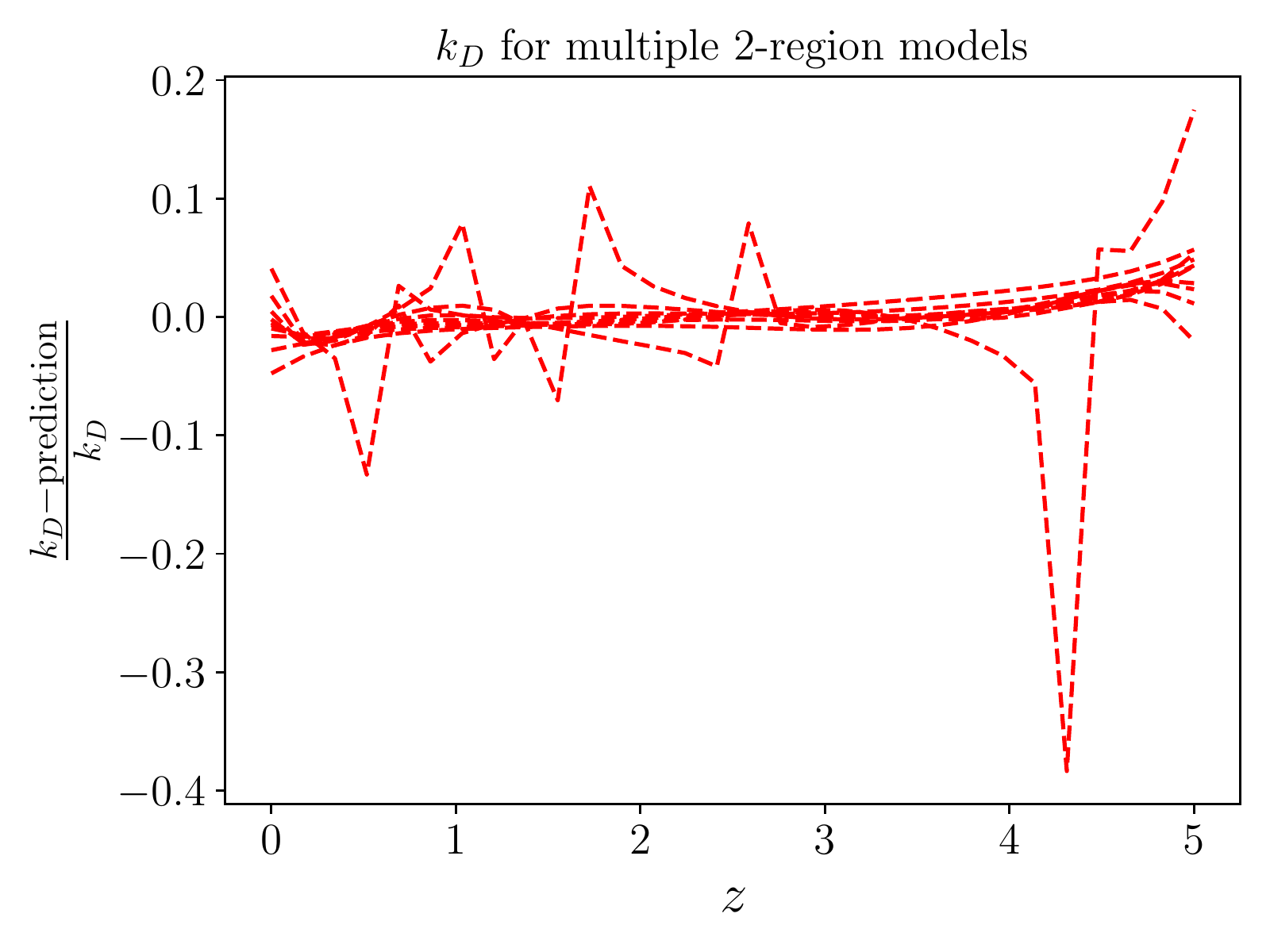}
	}	
	\caption{Curvature data $(z,f,10^{8}\cdot k_D)$ together with the most accurate expressions obtained with AI Feynman in the larger feature interval $z\in[0,5]$, $f\in[0.01,0.3]$. The spikes in the relative precision come from division by zero when $k_D$ changes sign.}
	\label{fig:kmult_zeq5}
\end{figure*}

\begin{figure*}
	\centering
		\subfigure[]{
		
	\includegraphics[scale = 0.5]{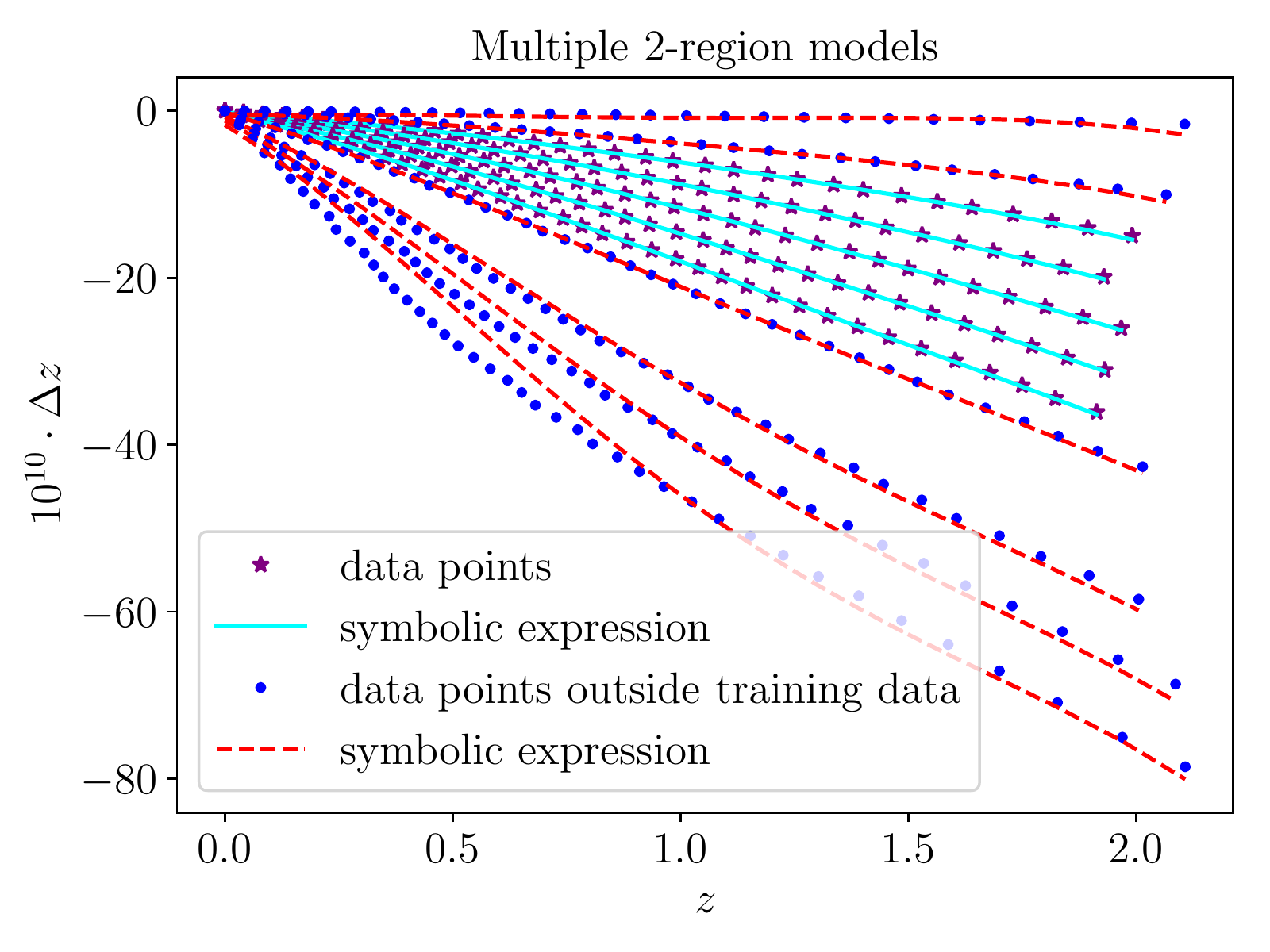}
}
		\subfigure[]{
	
	\includegraphics[scale = 0.5]{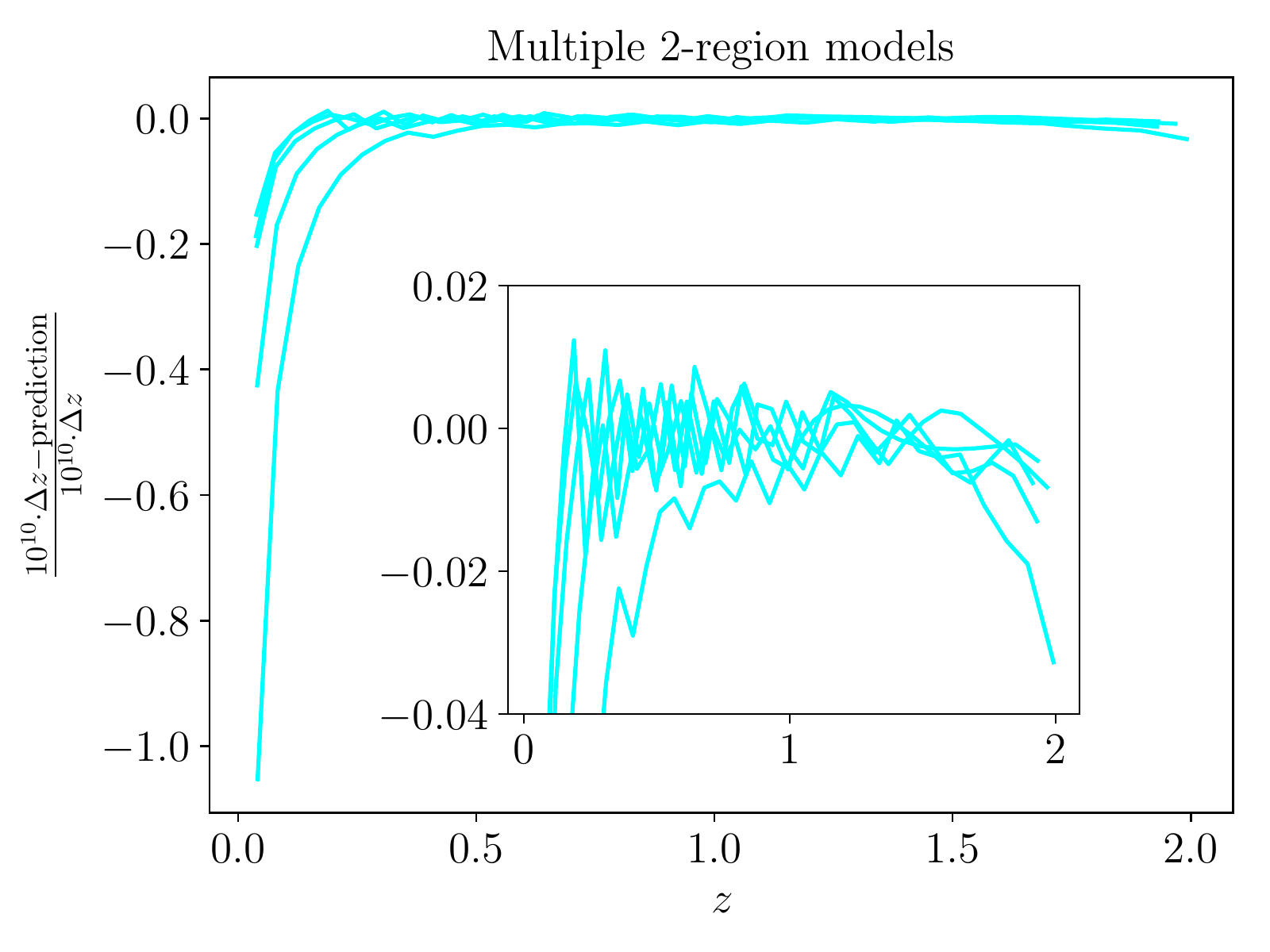}
}
	\caption{Symbolic expression for $(z, f, 10^{10}\cdot\Delta z)$ plotted together with data points. Stars and solid lines indicate data and symbolic expression, respectively, inside the feature regions used for generating data presented to AI Feynman. Dots and punctured lines indicate data and symbolic expression outside the region. The punctured lines are for $f = 0.01,0.06, 0.28,0.39,0.45,0.5$, in order of topmost to lowest lying line. The solid lines indicate data for $f = 0.1,0.1375,0.2125, 0.25$, from top to bottom. The relative accuracy of the symbolic expression is shown in the figure to the right.}
	\label{fig:Dz_multi}
\end{figure*}

\subsection{Multiple 2-region models: Symbolic regression}\label{subsec:regression_multi}
First, AI Feynman was used to attempt obtaining a symbolic expression for $(z,f,Q)$ in the data intervals used for the single 2-region model and the feature importance study. This proved very difficult with multiple attempts yielding expressions that all had inaccuracies at the order of 100\% over large parts of the feature intervals. Eventually, after minimizing the studied parameter intervals in steps, fairly accurate expressions were obtained by considering data of the type $(z,f,100\cdot Q)$ on the smaller intervals $z\in[0,2]$ and $f\in[0.1,0.25]$. The redshift interval was chosen not to be made smaller than this interval because it encapsulates the redshift area traced by the Square Kilometer Array (SKA) and thus contains the redshift interval where redshift drift measurements can be obtained with SKA \cite{SKA}. In addition, this redshift interval encompasses the main part of existing supernovae data which is particularly interesting in terms of constraining $Q$ using the redshift-distance relation. Thus, instead of making the redshift interval smaller beyond this interval, the $f$ interval was made successively smaller to attempt obtaining more accurate symbolic expressions. The regression task with $f\in[0.1,0.25]$ was still not wildly successful: Even after several attempts of using AI Feynman on the data, expressions that were accurate to sub-percent on the entire feature intervals were still not achieved. With three iterations it was, however, possible to obtain expressions with percent-precision for a large part of the feature region. The feature region was therefore not shrunken more after this. The most accurate results that were obtained are shown in figure \ref{fig:Qmult} which shows the results from the iterative approach of using AI Feynman on first the data $(z,f,100\cdot Q)$, then on $(z,f,100\cdot Q-f_1)$ and then $(z, f, 100\cdot(100\cdot Q-f_1-f_2))$, where $f_1$ and $f_2$ were the symbolic expressions obtained in the previous iteration. The resulting symbolic expressions are\footnote{AI Feynman presents expressions with a precision of 14 decimal places. To increase readability, a lower precision is shown here. Specifically, expressions are shown to a precision of five decimal places which should be sufficient for readers to verify the accuracy of the expressions while still allowing a reasonable readability.}
\begin{align}\label{eq:f1f2f3}
	\nonumber f_1 &= 3.58069\cdot\tan^{-1}\left( (f\cdot(z+\exp(z)))\right) \\\nonumber
	f_2 & = 0.089282\cdot z^4 + 1.19043\cdot z^3f- 0.51380\cdot z^3 \\\nonumber & - 15.87238\cdot z^2f^2 + 0.60670\cdot z^2f + 0.82042\cdot z^2 \\\nonumber &+ 4.19790\cdot zf^2+ 2.86470\cdot zf - 0.83058z\\\nonumber &+ 12\cdot f^2 +11.67448\cdot f^2 - 4.43136\cdot f + 0.38435\\
	f_3 &= -0.00002\cdot \exp(\exp(z)).
\end{align}
Note that these expressions represent $100\cdot Q$ and  not $Q$ itself.
\newline\indent
Figure \ref{fig:Qmult} also shows the relative accuracy of the symbolic expressions. As seen, the final fit has a percent-level precision for a significant part of the feature intervals but does become imprecise above 20\% in the considered feature region. If the symbolic expressions are used on feature intervals outside those used to create data sets for AI Feynman, the accuracy becomes very poor ({\em viz.} completely useless), very quickly.
\newline\indent
The expressions $f_1$ and $f_3$ were found to be the most accurate expressions identified by AI Feynman for the first and third iteration, respectively (and are therefore the only expressions shown here). From the second iteration, several expressions with similar accuracy were found, i.e. there were several choices for $f_2$ that looked equally promising  in terms of increasing the accuracy of the symbolic expression. Another example of $f_2$ identified by AI Feynman is
\begin{align}
f_2^{\rm new} = -0.02369+ z\frac{\sqrt{z}-1}{\pi},
\end{align}
but with this choice, no significant progress was made in terms of accuracy through a third iteration and the overall accuracy of the final symbolic expression was not as good as in the case presented in equation \ref{eq:f1f2f3}.
\newline\newline
The above results illustrate that the task of identifying an accurate symbolic expression for the kinematical backreaction for multiple 2-region models is ``difficult'', i.e. several attempts yielded complicated expressions that were only accurate on a smaller feature interval than what had originally been sought. It is therefore worth re-considering the choice of working with $Q$ rather than $R_D$; just as it was in section \ref{subsec:single} found that more accurate expressions were obtained for $Q$ than for $\Omega_{Q}$, it may be that more accurate expressions can be found for $R_D$. Results from applying AI Feynman to data of the form $(z, f, 10^8\cdot k_D)$ is shown in figure \ref{fig:kmult_zeq2}, where $k_D:=R_Da_D^2$ was chosen as the target for the regression task because $k_D$ is a constant in the FLRW limit, making it easy to compare the results with the FLRW limit. It was possible to use AI Feynman to obtain expressions for $k_D$ which were more accurate than what was achieved for $Q$. The most accurate fit obtained with AI Feynman is the polynomial
\begin{align}
	\nonumber 10^8\cdot k_D &=\\\nonumber
	& -0.26678\cdot z^6 + 1.75499\cdot z^5 - 2.05776\cdot z^4f \\\nonumber&
	- 4.32391\cdot z^4 + 9.42147\cdot z^3f + 4.97626\cdot z^3 \\\nonumber&
	- 15.87238\cdot z^2f^2 - 15.87188\cdot z^2f - 2.65310\cdot z^2\\\nonumber&
	 + 31.74477\cdot zf^2 + 24.06085\cdot zf + 0.30805\cdot z \\ &+ 21.31177\cdot f^2 - 3.21452\cdot f - 5.05345.
\end{align}
The relative accuracy of this expression is shown in figure \ref{fig:kmult_zeq2}. As seen, the expression found is overall quite accurate -- note that spikes in the relative accuracy comes from division by zero due to $k_D$ changing sign.
\newline\indent
When extrapolating outside the feature intervals used for training AI Feynman, the accuracy almost immediately becomes very poor. This is not surprising since this is a well-known quality of polynomials: They tend to fit data well within smaller intervals, but they are prone to overfitting and generally cannot be extrapolated outside the region they were originally fitted to. It is also worth noting that polynomial expressions for the curvature and kinematical backreaction is a generalization of the scaling relations discussed in the introduction. The scaling relations are, however, usually used as ``monomials'' rather than in polynomial versions with several terms. Nonetheless, since the polynomial expressions for $R_D = k_D/a_D^2$  can be related to the scaling relations, it becomes straightforward to obtain the corresponding expression for $Q$. This indicates that accurate polynomial expressions for $Q$ can also be obtained, making it a bit curious that the author was not successful in having the AI Feynman algorithm identify any such accurate polynomial expressions for $Q$\footnote{In relation to this comment it should be noted that polynomial expressions for $Q$ {\em were} obtained, including $f_2$ shown in the main text. In addition, a polynomial expression for $Q$ was also obtained as a fairly accurate version of $f_1$ i.e. during a ``first iteration''. This polynomial (not shown here) was third degree and obtained while the maximum polynomial degree permitted for AI Feynman was six. None of the polynomials obtained for $Q$ were as accurate as those found for $k_D$.}.
\newline\indent
The accurate fit obtained for $k_D$ encourages looking at the possibility of obtaining accurate fits on larger feature intervals. AI Feynman was therefore trained on data generated with the feature intervals used in section \ref{subsec:single}, i.e. $z\in[0,5]$ and $f\in[0.01,0.3]$. In this case the most accurate expression obtained is also a polynomial, namely
\begin{align}
	\nonumber 10^8\cdot k_D &=\\\nonumber & -0.00219 \cdot z^6 + 0.03768\cdot z^5 f + 0.03168\cdot z^5\\\nonumber & - 0.64965\cdot z^4 f^2 - 0.43309\cdot z^4 f - 0.17684\cdot z^4\\\nonumber &  + 8.84545\cdot z^3 f^2 + 1.52360\cdot z^3 f+ 0.49810\cdot z^3\\\nonumber &  + 8.09975\cdot z^2f^3 - 45.06587\cdot z^2f^2 - 1.95429\cdot z^2f\\\nonumber & - 0.78006\cdot z^2 - 40.49877\cdot zf^3 + 80.42749\cdot zf^2\\\nonumber & + 9.97251\cdot zf + 0.67266\cdot z + 67.56565\cdot f^3\\ & - 25.90655\cdot f^2 + 7.02894\cdot f - 5.75576.
\end{align}
A comparison of this expression with data is shown in figure \ref{fig:kmult_zeq5}. Again, the expression is accurate to percent-order or better for most of the considered feature region. (And again, there are spikes in the accuracy curves that come from $k_D$ crossing the value zero.)
\newline\newline
We now move on to look at the redshift drift. AI Feynman was first presented with data sets of the type $(Q_{D,100}, f,10^{10}\cdot\Delta z)$ with the feature intervals $z\in[0,5]$ and $f\in[0.01,0.3]$, but again it proved difficult to obtain accurate symbolic expressions for these data sets. As for the data for $Q$, accurate expressions were eventually obtained for a data set of the type $(z, f, 10^{10}\cdot \Delta z)$, using the smaller intervals $z\in[0,2]$ and $f\in[0.1,0.25]$. Figure \ref{fig:Dz_multi} shows data points together with the most accurate symbolic expression obtained. The only symbolic expression found that was accurate on almost the entire feature interval was the polynomial
\begin{align}
\nonumber 10^{10}\Delta  z & =\\
&\nonumber -10^{-5}\cdot z^6 - 0.05383\cdot z^5 f - 0.16136\cdot z^5 \\\nonumber &- 51.42798\cdot z^4f^2 + 16.24516\cdot z^4f - 0.71557\cdot z^4 \\\nonumber &- 1.41156\cdot z^3f^3 + 207.32051\cdot z^3 f^2 - 58.40215\cdot z^3 f\\\nonumber & + 4.05026\cdot z^3 - 0.01453\cdot z^2f^4
 + 4.26328\cdot z^2f^3 \\\nonumber &- 191.11012\cdot z^2f^2 + 41.45695\cdot z^2 f - 5.05871\cdot z^2 \\\nonumber &- 7\cdot 10^{-5}\cdot zf^5 + 0.02924\cdot z f^4 - 2.61453\cdot zf^3\\\nonumber & - 30.04826\cdot zf^2 - 56.13623\cdot zf + 2.28806\cdot z \\\nonumber+ & 7\cdot 10^{-5}\cdot f^5 - 0.00896\cdot f^4 - 0.20814\cdot f^3\\ & - 16.98247\cdot f^2 + 6.61763\cdot f - 0.72982.
\end{align}
Figure \ref{fig:Dz_multi} also shows the relative deviation between $10^{10}\cdot \Delta z$ and the prediction of the symbolic expression shown above. As seen, on the main part of the feature region, the predictions by the symbolic expression have an error around or below 1\%.
\newline\indent
As illustrated in figure \ref{fig:Dz_multi}, the symbolic expression was also compared to data points outside the feature intervals of the data presented to the AI Feynman algorithm. As seen, the model actually extrapolates fairly well a bit outside the $f$-interval it was developed on. It does not, however, extrapolate well to significantly larger values of the redshift, where it quickly becomes highly inaccurate.
\newline\indent
It is lastly noted that some of the expressions found by AI Feynman which were simpler (i.e. with fewer terms but non-polynomial) than the one shown above had percent-level accuracy on large parts of the feature intervals as well, but they were not quite as accurate as the polynomial expression which is why only the polynomial expression is shown here. Simplicity and accuracy are both important qualities of symbolic expressions, and how these two qualities should be weighed against each other is not clear and certainly depends on the goal with the expressions. For instance, if one wishes to study a possible physical justification behind symbolic expressions, it may be prudent to weigh simplicity higher. Here, simplicity was not considered by the author\footnote{But note that simplicity is rewarded by the AI Feynman algorithm which means that simplicity was still indirectly used to select symbolic expressions.} and the selection of symbolic expressions focused only on accuracy, simply because the main point here is that it is actually possible to obtain accurate expressions.

\section{Discussion and conclusion} \label{sec:conclusion}
Symbolic expressions for the kinematical backreaction, spatially averaged spatial curvature and the redshift drift in 2-region models were obtained through symbolic regression based on the publicly available AI Feynman algorithm. It proved difficult to achieve expressions with sub-percent accuracy for the kinematical backreaction in terms of the (mean) redshift and the model parameter $f$. It was much easier to obtain accurate symbolic expressions for the curvature and the redshift drift. Indeed, several expressions with around 1 \% or sub-percent accuracy for $k_D$ and $\left\langle \delta z\right\rangle $ were obtained, perhaps the most interesting being an expression for $\left\langle \delta z\right\rangle  = \delta \left\langle z\right\rangle  + \Delta z$. In this expression, $\delta\left\langle z\right\rangle $ represents the naive expression equivalent to the FLRW limit, and a symbolic expression for $\Delta z$  was found with AI Feynman. Regarding the expressions obtained for $k_D$ in terms of the redshift and model parameter $f$, it is worth noting that the accurate expressions obtained were all polynomials. This is interesting because it highlights that the models obtained by symbolic regression should generally be expected to be phenomenological, but also because the expression can be considered a generalization of the scaling solutions for backreaction studied in existing literature.
\newline\indent
The fact that different, roughly equally accurate, expressions were obtained for $Q(z,f)$, $k_D(z,f)$ as well as $\Delta z$ as a function of either $Q$ or $z$ together with $f$ is another  reminder that the expressions themselves were not obtained through theoretical considerations and therefore do not necessarily represent theoretical insight but instead represent phenomenological models. The lack of theoretical insight is the big downside with symbolic regression and, indeed, machine learning in general. However, the obtained expressions are still useful; although the expressions have no significant theoretical underpinning, they are still correct (phenomenological) reflections of the relationships between the different variables/features. This means that it is still valid to use the expressions for e.g. constraining model parameters with observational data. This point is the main motivation for the work presented here. In addition, the concept of cosmic backreaction has been a significant part of the cosmological literature for over two decades now, but very little is still known about under what circumstances $Q$ will be non-negligible and how $Q$ can be parameterized in terms of $z$. Similarly, although the redshift drift is clearly an important future observable, it has so far not been possible to understand how redshift drift is related to spatial averages in a general spacetime. Combining theoretical work with symbolic regression or other types of machine learning may pave the way forward. Note for instance that \emph{if} a relation between spatial averages and mean redshift drift exists under certain restrictions e.g. similarly to equation \ref{eq:av_DA}, then an exhaustive symbolic regression algorithm {\em will} be able to find it, although it may require much fine-tuning of the hyperparameters, selecting among algorithms, and patience. If such a physically justifiable relation between spatial averages and mean observations is obtained through symbolic regression, it is identifiable since it must be possible to extrapolate the symbolic expression e.g. beyond the feature intervals used for the regression as well as to other models. By e.g. studying under which model assumptions the obtained expression is valid, theoretical insight regarding the underlying physical justification for the expression can be revealed.
\newline\indent
The results presented here were obtained using AI Feynman which is aimed at solving regression tasks within a broad range of fields within physics. It is expected that more accurate symbolic expressions for backreaction and redshift drift can be achieved by using symbolic regression algorithms tuned for this specific task. The fact that accurate symbolic expressions could be obtained here for a significant parameter region using AI Feynman without modifying the algorithm or fine-tuning its hyperparameters (listed in table II of \cite{AIFeynman_1}) gives reason to be optimistic that more complex backreaction evolution can also be described using symbolic regression, at least if one tunes the algorithms appropriately. It therefore seems reasonable to be optimistic that symbolic regression performed on data from more realistic models can yield accurate symbolic expressions for both $Q$ and $\delta z$ over long ranges of redshift and hence diverse types of observations. This is especially the case since there seems to be no reason to expect that more realistic and hence complicated models should necessarily imply more complex cosmic backreaction. Since cosmic backreaction vanishes or becomes very small under certain constraints in even some of the most realistic cosmological models currently available including relativistic simulations (see e.g. \cite{HayleyBackreaction, gevolution}), it is in fact not implausible that backreaction evolution is simpler to model for more realistic models than for the models considered here. It is, on the other hand, of course also possible that new obstacles will turn up when considering other models. One possible obstacle that could require special modifications of the algorithm is if one attempts obtaining a symbolic expression for $\Delta z$ in terms of $Q$ in a case where $Q$ is not monotonic in $z$.
\newline\newline
Lastly, it must be stressed that the symbolic expressions obtained here are only valid for the studied 2-region models in the studied parameter intervals. Since there is no reason to expect that these models reflect backreaction realistically, the expressions obtained here should not be used to attempt \emph{realistic} parameter constraining with real data. The results are still useful in terms of parameter constraints for e.g. proof-of-principle studies. Future studies will focus on utilizing the approach presented here together with more realistic backreaction and redshift drift data in order to gain a more general (and realistic) idea of how these are related to volume averaged quantities. It could also be interesting to look at more sophisticated toy-models such as those of \cite{multiscale1, multiscale2, simple_timescape} which generalize the simple model studied here.

	\vspace{6pt} 
	\begin{acknowledgments}
		 The author thanks the anonymous referees for their careful comments and good suggestions which have significantly improved the presentation of the work. During the final stages of the review process, the author transitioned from being funded by the Carlsberg foundation to being funded by VILLUM FONDEN (grant VIL53032).
		
	\end{acknowledgments}



\end{document}